%% file: JET_GENE_compare.tex
\documentclass[12pt]{iopart}

\usepackage{graphics}
\usepackage{graphicx}
\usepackage{enumerate}
\begin{document}

\title[Ion temp. profile stiffness: non-lin. gyrokinetic simu. and comparison with exp.]{Ion temperature profile stiffness: non-linear gyrokinetic simulations and comparison with experiment}

\author{J.~Citrin$^1$, F.~Jenko$^2$, P.~Mantica$^3$, D.~Told$^2$, C.~Bourdelle$^4$, R.~Dumont$^4$, J.~Garcia$^4$, J.W.~Haverkort$^{1,5}$, G.M.D.~Hogeweij$^1$, T. Johnson$^6$,  M.J.~Pueschel$^7$, and JET-EFDA contributors$^*$}

\address{JET-EFDA, Culham Science Centre, Abingdon, OX14 3DB, UK}
\address{$^1$FOM Institute DIFFER -- Dutch Institute for Fundamental Energy Research - Association EURATOM-FOM, Nieuwegein, The Netherlands}
\address{$^2$Max Planck Institute for Plasma Physics, EURATOM Association, Boltzmannstr. 2, 85748 Garching, Germany}
\address{$^3$Istituto di Fisica del Plasma ``P. Caldirola'', Associazione Euratom-ENEA-CNR, Milano, Italy}
\address{$^4$CEA, IRFM, F-13108 Saint Paul Lez Durance, France}
\address{$^5$Centrum Wiskunde \& Informatica (CWI), PO Box 94079, Amsterdam, The Netherlands}
\address{$^6$Euratom-VR Association, EES, KTH, Stockholm, Sweden}
\address{$^7$University of Wisconsin-Madison, Madison, Wisconsin 53706, USA}
\address{\vspace{2mm} $^*$See the Appendix of F. Romanelli et al., Proceedings of the 24th IAEA Fusion Energy Conference 2012, San Diego, USA}
\ead{J.Citrin@differ.nl}

\begin{abstract}
Recent experimental observations at JET show evidence of reduced ion temperature profile stiffness, hypothesised to be due to concomitant low magnetic shear ($\hat{s}$) and significant toroidal rotational flow shear. Non-linear gyrokinetic simulations are performed, aiming to investigate the physical mechanism behind the observations. A comprehensive set of simulations are carried out, comparing the impact on the ion heat flux of various parameters that differ within the data-set. These parameters include $q$, $\hat{s}$, rotation, effect of rotation on the magnetohydrodynamic (MHD) equilibrium, $R/L_n$, $\beta_e$, $Z_{eff}$, and the fast particle content. The effect of toroidal flow shear itself is not predicted by the simulations to lead to a significant reduction in ion heat flux, due both to an insufficient magnitude of flow shear and significant parallel velocity gradient destabilisation. It is however found that non-linear electromagnetic effects due to both thermal and fast-particle pressure gradients, even at low $\beta_e$, can significantly reduce the profile stiffness. A total of five discharges are examined, at both inner and outer radii. For all cases studied, the simulated and experimental ion heat flux values agree within reasonable variations of input parameters around the experimental uncertainties.  %However, no such reasonable agreement is obtained for the discharge with the highest logarithmic ion temperature gradient at inner radii; the simulated stiffness level is higher than experimentally observed for this regime. The observed low stiffness thus remains unexplained when assuming pure toroidal rotation. A further option for explaining the flux reduction which remains to be explored is anomalous poloidal rotation. The degree of anomalous poloidal rotation necessary to reduce the flux level to the observed value is estimated.
\end{abstract}

\pacs{52.30.Gz, 52.35.Ra, 52.55.Fa, 52.65.Tt}

%\newpage
\input{sect1}

\input{sect2}
\input{sect3}

\input{sect4}
\input{sect5}
\input{sect6}
\input{sect7}

\section*{References}

\bibliographystyle{unsrt}
\bibliography{rotationbib}

%\input{figssect2}
%\clearpage
%\input{figssect3}
%\clearpage
%\input{figssect4}
%\clearpage

%\input{alltabs}
%\input{allfigs}

\end{document}

%% file: sect1.tex
\section{Introduction}
\label{sec:intro}

It is well established that one of the primary limitations of tokamak energy confinement is ion-Larmor-radius-scale turbulent transport driven by background pressure gradients~\cite{ITER2}. The ion-temperature-gradient (ITG) instability in particular has been long identified as a ubiquitous unstable mode in tokamak plasmas~\cite{roma89,bigl89,guo93}, and is primarily responsible for ion heat losses. The instability saturates in a non-linear state in conjunction with non-linearly excited zonal-flows, forming a self-organised turbulent system which sets the transport fluxes~\cite{diam05}. 

In addition to self-organised zonal-flows, the application of external flow shear is predicted to suppress turbulence through two broad mechanisms: decorrelation of the turbulent structures in the non-linear phase, once the shearing rate is comparable with or exceeds the inverse non-linear autocorrelation time; and suppression of the driving linear modes by continuously shifting the mode from the most unstable spatial scale to nearby, more stable spatial scales~\cite{burr97,terr00}. Flow shear has been observed experimentally to be correlated with ion temperature tokamak transport barriers~\cite{wagn07,crom05,devr09}. Non-linear gyrofluid simulations with adiabatic electrons have predicted turbulence quenching above $\gamma_E/\gamma_{max}=1\pm0.3$~\cite{walt94,walt95}, where for purely toroidal rotation the normalised $E{\times}B$ shear rate $\gamma_E{\equiv}\frac{r}{q}\frac{d\Omega}{dr}/(\frac{v_{th}}{R})$, and $\gamma_{max}$ is the maximum linear growth rate in the absence of rotation.  Later gyrokinetic simulations, including cases with kinetic electrons, predicted that quenching occurs at somewhat higher (but similar) flow shear compared to the earlier gyrofluid simulations, at $\gamma_E/\gamma_{max}=2\pm0.5$~\cite{walt02,kins05c,roac09}. According to results in Ref.\cite{kins05c}, this quench behaviour is independent of the adiabatic or kinetic electron assumption. However, when including kinetic electrons, the trapped electron drive tends to raise the instability growth rates even for ITG turbulence. Therefore, when including kinetic electrons (which is more realistic) the resultant higher $\gamma_{max}$ necessitates a higher value of $\gamma_E$ compared with the adiabatic electron case to reach a similar $\gamma_{E}/\gamma_{max}$ ratio and quench the turbulence. The seeming robustness of the transport quench has motivated formulations of effective growth rate reduction due to the flow shear in the mixing length rule of quasilinear transport models such as GLF23 and TGLF~\cite{walt97,kins08,stae13}.

The abovementioned quench results were obtained in simulations which did not include a self-consistent parallel velocity gradient (PVG) in the system, which can be destabilising~\cite{catt73,newt10}. When PVG destabilisation is included, simulations have shown that it can limit the transport quench~\cite{walt94,dimi01,kins05c}. It is thus important to incorporate the effects of PVG destabilisation in transport models when comparing modelling predictions with experiments that have significant flow shear. For pure toroidal rotation, the degree of the PVG destabilisation depends on the magnetic geometry through the ratio $q/\epsilon$. In the pure toroidal rotation case, $\gamma_p=\frac{q}{\epsilon}\gamma_E$, where $\gamma_p$ is the PVG shear rate, and $\epsilon{\equiv}r/R$. 

%Experimental results of flow shear stabilisation of core turbulence are not systematic.

The impact of the rotational flow shear on global confinement in experiments has been observed to vary. While improved global confinement in hybrid scenarios due to flow shear has been observed at DIII-D~\cite{poli08}, experiments involving JET standard H-modes showed a lack of significant variation in global confinement even when significantly varying the rotation profiles and maintaining constant total power~\cite{vers11}.% These differences in observed behaviour are likely due to the dependence of the flow shear induced turbulence reduction on the magnetic geometry through $q/\epsilon$ and magnetic shear~\cite{mant11}, as well as on the details of the power deposition profiles in the experiments.

It has been recently observed in dedicated experiments at JET that ion temperature profile stiffness can be reduced at low normalised radii ($r/a<0.5$), in disagreement with non-linear ITG turbulence modelling~\cite{mant09,mant11,mant11b}. This has been hypothesised to be related to the correlation between low magnetic shear ($\hat{s}$) and increased flow shear in the low stiffness discharges. The term `stiffness' is defined here as the local gradient of the gyro-Bohm normalised ion heat flux with respect to $R/L_{Ti}$ (normalised inverse ion temperature gradient length). The observations concentrated on $\rho=0.33$ and $\rho=0.64$ (where $\rho$ is the normalised toroidal flux coordinate). At $\rho=0.33$, the stiffness is observed to transit from high to low when the flow shear was increased. However, at $\rho=0.64$, stiffness is observed to be high irrespective of flow shear. A previous non-linear gyrokinetic study based on the recent JET discharges at $\rho=0.33$, as detailed in Ref.\cite{mant11}, reported only an ITG threshold shift with rotation, as opposed to a decrease in stiffness as observed. Additional observations made in Refs.~\cite{mant09,mant11,mant11b} pertinent to this work are as follows: at low rotation at $\rho=0.33$, the observed stiffness level is higher than the gyrokinetic simulation predictions; furthermore, the observed ITG threshold is lower than the non-linear gyrokinetic prediction, questioning the manifestation of the Dimits shift~\cite{dimi00} predicted by non-linear simulations.

\begin{figure}[htbp]
	\centering
		\includegraphics[scale=0.7]{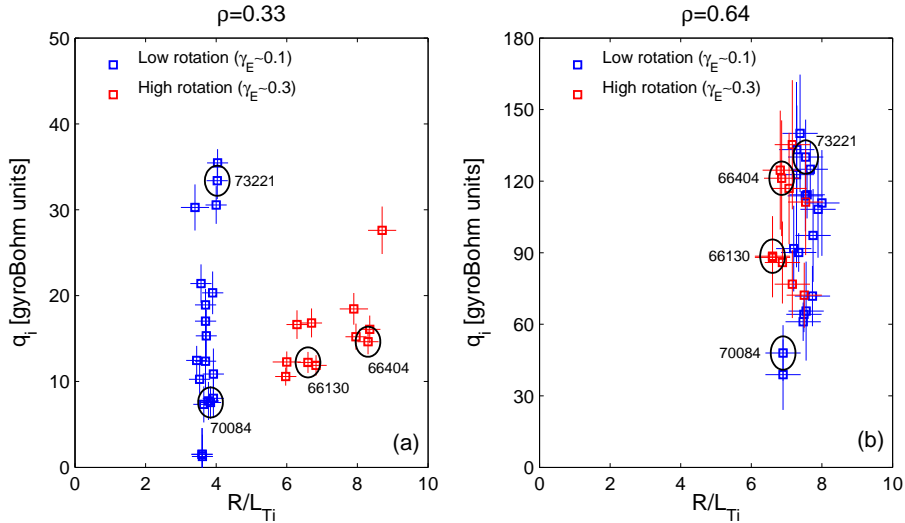}
		\caption{Partial reproduction of data presented in Ref.\cite{mant11} displaying the separation between high and low stiffness regimes at $\rho=0.33$ (a) for discharges with low and high rotation respectively. At $\rho=0.64$ (b) no significant separation of the stiffness behaviour is evident. The heat fluxes are in gyroBohm normalised units, $q_{GB}=T_i^{2.5}n_im_i^{0.5}/e^2B^2R^2$. The specific discharges studied in this paper have been circled.} %A representative `low stiffness' $R/L_T(q_i)$ curve has been produced by a simple linear fit to the high rotation low stiffness data. }
	\label{fig:figure1}
\end{figure}

In this paper, we extend this previous work and investigate whether the experimental observations can be understood through gyrokinetic modelling. Understanding these effects could allow the identification of a potential actuator for core $T_i$ control. As opposed to the previous simulation work, we include numerical geometry, electromagnetic effects, fast particles, parallel velocity gradient destabilisation, and explore the impact of reasonable variations in input parameters (from the experimental data) such as $\hat{s}$, $q$, and $R/L_n$. For the analysis, linear and non-linear simulations are carried out with the \textsc{Gene} code~\cite{jenk00}. Four JET discharges (with the previous carbon wall) were selected: 70084, 66130, 66404, and 73221. Discharges 66130 and 66404 are situated on the `high-rotation, low-stiffness branch' at $\rho=0.33$ seen in Fig.~1 in Ref.\cite{mant11}, and partially reproduced here for convenience in Fig.~\ref{fig:figure1}. We note that discharge 66404 has also been analyzed in Ref.~\cite{ryte11}, where the possibility of increased critical threshold in conjunction with the lowered stiffness is not ruled out. Discharge 70084 is a low flux, low rotation discharge selected to provide a data point near the turbulence threshold. Discharge 73221 is a high flux, low rotation discharge situated on the `low-rotation, high-stiffness branch' at $\rho=0.33$, as shown in Fig.~\ref{fig:figure1}. The specific questions which we investigate are the following:

\begin{enumerate}[(1)]
\item Is the experimentally observed stiffness reduction for the high-rotation discharges at $\rho=0.33$ consistent with gyrokinetic non-linear simulation predictions? Which plasma parameters have the highest impact on the stiffness level for ITG turbulence? Is there sufficient leeway in the plasma parameters due to uncertainties such that the experimental observations and non-linear simulation predictions can be reconciled?
\item Can the seeming high stiffness in the `low-rotation, high-stiffness' branch at $\rho=0.33$ be reconciled with the non-linear simulations, given reasonable uncertainties in plasma parameters?
\item At $\rho=0.64$, is the \textit{lack} of experimentally observed stiffness reduction for the high-rotation discharges consistent with gyrokinetic non-linear simulation predictions?
\item Can the experimentally extrapolated turbulence threshold be reconciled with the non-linear turbulence threshold including the Dimits shift, given reasonable uncertainties in the plasma parameters?
\end{enumerate}

The discharges were reanalysed with the \textsc{cronos} suite of integrated modelling codes~\cite{arta10} to identify any differences in parameters apart from rotation and $R/L_{Ti}$ within the chosen discharge set - such as $T_e/T_i$, $R/L_n$, $\beta_e$, $q$, $\hat{s}$, and fast particle content - that may lead to the observed differences in ion heat flux and $R/L_{Ti}$. The sensitivity of the ion heat flux and stiffness to each of these parameters was tested with \textsc{Gene} in dedicated $R/L_{Ti}$ scans. Finally, complete simulations - i.e. collisional, electromagnetic, multi-species, and with rotation - were carried out at both $\rho=0.33$ and $\rho=0.64$. %to determine whether the gyrokinetic simulations can reproduce the experimental fluxes. %We stress that one of the benefits of our set of parameter scans is that it consists of numerous cases situated in the vicinity of the turbulence threshold, at experimentally relevant fluxes. We foresee that these simulations will prove particularly useful in validating reduced transport models.

The rest of the paper is organised as follows. In section~\ref{sec:gene} the \textsc{Gene} gyrokinetic code is briefly reviewed, as are the base parameters of the simulated discharges. Section~\ref{sec:lowshear} discusses the stiffness sensitivity study at $\rho=0.33$. Section~\ref{subsec:all} shows the full comparison between the ion heat flux measurements and gyrokinetic predictions at $\rho=0.33$. In section~\ref{sec:highshear} the same comparison at $\rho=0.64$ is shown. Conclusions are presented in section~\ref{sec:discuss}.

%% file: sect2.tex
\section{GENE simulations and discharge parameters}
\label{sec:gene}

\textsc{Gene} solves the gyrokinetic Vlasov equation, coupled self-consistently to Maxwell's equations, within a ${\delta}f$ formulation~\cite{briz07}. Computational efficiency is gained by solving in field line coordinates. $x$ is the radial coordinate, $z$ is the (poloidal) coordinate along the field line, and $y$ is the binormal coordinate. Both an analytical circular geometry model (derived in Ref.\cite{lapi09}) as well as a numerical geometry were used in this work. The circular geometry model avoids the order $\epsilon=a/R$ inconsistency present in the often applied $s-\alpha$ model, but does not include a Shafranov shift. For the numerical geometry, the \textsc{finesse} code was used to solve the extended Grad-Shafranov equation including toroidal rotation~\cite{beli02}. All simulations carried out were local, which is justified since $1/\rho^*{\sim}500$ for the range of plasma parameters studied here~\cite{cand04,mcmi10}. Both linear and non-linear simulations were performed. In the linear mode, an eigenvalue solver was used to compute multiple modes for each point in parameter space~\cite{kamm08,merz12}. In the presence of rotation, when no time-independent eigenmodes can form, a complementary initial value solver was used.

Four discharges from the data-set presented in Ref.\cite{mant11} were analyzed at $\rho=0.33$ and $\rho=0.64$, where $\rho$ is the normalised toroidal flux coordinate. The discharges are 70084, 66130, 66404, and 73221. Discharge 70084 corresponds to a representative low rotation, low flux discharge. 66130 and 66404 are discharges further up on the `high rotation, decreased stiffness' curve as seen in Fig.~\ref{fig:figure1}. 73221 is a high flux, low rotation discharge situated on the `low-rotation, high-stiffness branch' at $\rho=0.33$, as shown in Fig.~\ref{fig:figure1}. The kinetic profiles of the four discharges were spline fitted and interpretative runs were carried out with the \textsc{cronos} integrated modelling suite of codes~\cite{arta10} for the equilibrium calculations and $q$-profile calculations. The kinetic profiles were then averaged over $1$~s centered around $10/10/7/7.5$~s respectively for calculations of the gradient lengths and other quantities such as $\beta_e$. The parameters are shown in tables \ref{tab:summary1}-\ref{tab:summary2}. Discharge 73221 was only analysed at $\rho=0.33$, for the investigation of the seemingly high stiffness of the low rotation branch. The ${\langle}Z_{eff}{\rangle}$ values correspond to Bremsstrahlung measurements. Since the precise $Z_{eff}$ profiles are not known, the sensitivity of the transport predictions to the range of reasonable $Z_{eff}$ at $\rho=0.33$ is explored in section~\ref{subsec:all}. $\nu^*$ is the normalised collisionality: $\nu^*{\equiv}\nu_{ei}\frac{qR}{\epsilon^{1.5}v_{te}}$, with $\epsilon=a/R$ and $v_{te}=\sqrt{\frac{T_e}{m_e}}$. Note that the data presented in table~\ref{tab:summary2} was processed \textit{separately and independently} from the values quoted in Ref.\cite{mant09,mant11} and shown in Fig.~\ref{fig:figure1}. The $R/L_{Ti}$ values in table \ref{tab:summary2} and Fig.~\ref{fig:figure1} agree within error bars.

\begin{table}
\small
\centering
\caption{\footnotesize Discharge dimensional parameters. The values are averaged between 9.5-10.5 s for discharges
70084 and 66130, between 6.5-7.5 s for discharge 66404, and between 7-8 s for discharge 73221. Quoted errors are statistical, and do not include possible systematic errors.}
%\vspace{0.15cm}
\tabcolsep=0.11cm
\scalebox{0.7}{\begin{tabular}{c|c|c|c|c|c}
\label{tab:summary1}
Shot no.@location & $B$ [T]& $I_p$ [MA] & $T_i$ [keV] & $T_e$ [keV] & $n_e$ [$10^{19}$~m$^{-3}$] \\
\hline
70084@$\rho=0.33$ & $3.5$ & $1.8$ & $2.01\pm0.02$ & $2.16\pm0.1$ & $2.6\pm0.2$  \\
66130@$\rho=0.33$ & $3.1$ & $1.5$ & $2.58\pm0.04$ & $3.3\pm0.3$ & $3.37\pm0.24$  \\
66404@$\rho=0.33$ & $3.5$ & $1.8$ & $3.1\pm0.13$ & $3.61\pm0.08$ & $2.3\pm0.1$ \\
73221@$\rho=0.33$ & $3.5$ & $1.8$ & $1.84\pm0.04$ & $2.44\pm0.03$ & $2.45\pm0.16$ \\
\hline
\hline
70084@$\rho=0.64$ & $3.5$ & $1.8$ & $1.08\pm0.02$ & $1.27\pm0.05$ & $2.3\pm0.3$ \\
66130@$\rho=0.64$ & $3.1$ & $1.5$ & $1.38\pm0.03$ & $1.5\pm0.24$ & $2.8\pm0.3$  \\
66404@$\rho=0.64$ & $3.5$ & $1.8$ & $1.34\pm0.08$ & $1.66\pm0.14$ & $1.51\pm0.07$ \\
\hline
\end{tabular}}
\end{table}

\begin{table}
\small
\centering
\caption{\footnotesize Discharge dimensionless parameters. The $\hat{s}$ and $q$ values are calculated by \textsc{cronos} interpretative simulations, assuming neoclassical diffusion. The values are averaged between 9.5-10.5 s for discharges 70084 and 66130, between 6.5-7.5 s for discharge 66404, and between 7-8 s for discharge 73221.}
%\vspace{0.15cm}
\tabcolsep=0.11cm
\scalebox{0.7}{\begin{tabular}{c|c|c|c|c|c|c|c|c|c|c}
\label{tab:summary2}
Shot no.@location & $\hat{s}$ & $q$ & $T_e/T_i$ & $R/L_{Ti}$ & $R/L_{Te}$ & $R/L_{ne}$ & $\beta_e$ [\%] & $\nu^*$ & ${\langle}Z_{eff}{\rangle}$ & $M [v_{tor}/c_s]$ \\
\hline
70084@$\rho=0.33$ & $0.7$ & $1.7$ & $1.08\pm0.04$ & $3.5\pm0.5$ & $3.8\pm0.6$ & $1.4\pm0.4$ & $0.19\pm0.01$ & $0.07$ & $2.2\pm0.1$ & $0.09$  \\
66130@$\rho=0.33$ & $0.7$ & $1.8$ & $1.25\pm0.13$ & $6\pm0.4$   & $6.5\pm1$   & $2.4\pm1$   & $0.46\pm0.09$ & $0.04$ & $1.8\pm0.1$ & $0.31$  \\
66404@$\rho=0.33$ & $0.4$ & $1.8$ & $1.14\pm0.06$ & $8.6\pm0.9$ & $5.5\pm0.8$ & $3.8\pm0.4$ & $0.35\pm0.07$ & $0.02$ & $2.2\pm0.1$ & $0.19$  \\
73221@$\rho=0.33$ & $0.7$ & $1.5$ & $1.33\pm0.02$ & $3.8\pm0.4$ & $5.4\pm0.2$ & $2.8\pm0.3$ & $0.2\pm0.02$  & $0.055$ & $2.2\pm0.1$ & $0.07$  \\
%73224@$\rho=0.33$ & $0.5$ & $1.7$ & $1.0\pm0.01$ & $3.8\pm0.4$ & $6.8\pm0.1$ & $1.3\pm0.1$ & $0.33\pm0.004$  & $0.038$ & $2.3\pm0.1$ & $0.2$  \\
\hline
\hline
70084@$\rho=0.64$ & $1.3$ & $3$ & $1.18\pm0.05$ & $7.2\pm0.2$ & $6.4\pm1$ & $1.8\pm0.8$ & $0.096\pm0.01$ & $0.16$ & $2.2\pm0.1$ & $0.03$  \\
66130@$\rho=0.64$ & $1.5$ & $3.5$ & $1.1\pm0.2$ & $6.8\pm0.3$ & $8.5\pm3$ & $1.8\pm1.4$ & $0.18\pm0.04$ & $0.1$ & $1.8\pm0.1$ & $0.23$  \\
66404@$\rho=0.64$ & $1.4$ & $2.9$ & $1.23\pm0.13$ & $6.9\pm0.4$ & $10\pm1.6$ & $2.1\pm0.9$ & $0.08\pm0.01$ & $0.05$ & $2.2\pm0.1$ & $0.15$  \\
%73221@$\rho=0.64$ & $0.7$ & $1.5$ & $1.33\pm0.02$ & $8.4\pm0.3$ & $5.4\pm0.2$ & $2.8\pm0.3$ & $0.2\pm0.02$  & $0.055$ & $2.2\pm0.1$ & $0.07$ \\
\hline
\end{tabular}}
\end{table}

\begin{figure}[htbp]
	\centering
		\includegraphics[scale=0.45]{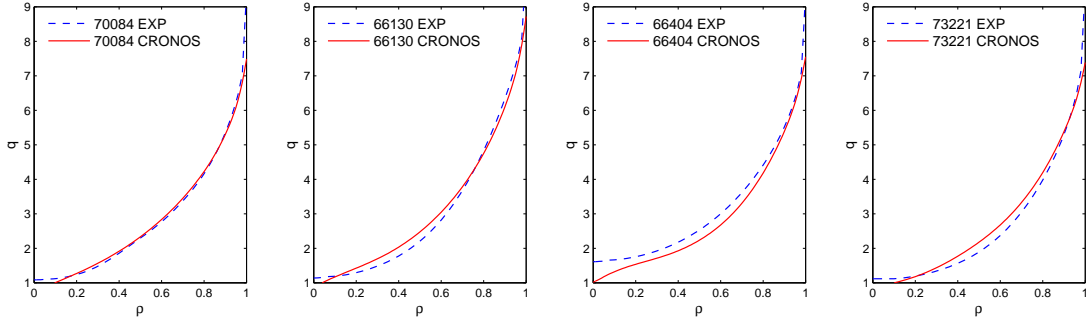}
		\caption{Comparison between \textsc{cronos} interpretative simulation $q$-profiles and experimental $q$-profiles. The profiles are averaged between 9.5-10.5~s for discharges 70084 and 66130, between 6.5-7.5~s for discharge 66404, and between 7-8 s for discharge 73221.}
	\label{fig:figure2}
\end{figure}

%\begin{table}
%\small
%\centering
%\caption{\footnotesize Motional Stark effect (66130 and 66404) and polarimetry (70084) $q$-profile measurements}
%%\vspace{0.15cm}
%\tabcolsep=0.11cm
%\scalebox{0.6}{\begin{tabular}{c|c|c}
%\label{tab:summary1}
%Shot no.@location & $\hat{s}$ & $q$ \\
%\hline
%70084@$\rho=0.33$  & $0.75\pm0.02$ & $1.57\pm0.03$\\
%66130@$\rho=0.33$ & $0.56\pm0.01$ & $1.58\pm0.07$ \\
%66404@$\rho=0.33$ & $0.37\pm0.04$ & $2\pm0.2$ \\
%\hline
%\hline
%70084@$\rho=0.64$ & $1.29\pm0.04$ & $3\pm0.03$  \\
%66130@$\rho=0.64$ & $1.67\pm0.07$ & $3.15\pm0.08$ \\
%66404@$\rho=0.64$ & $1.28\pm0.04$ & $3.3\pm0.1$ \\
%\hline
%\end{tabular}}
%\end{table}

%10^-4 6.2  3.7  2  14.5  9.07  6.09
The agreement between the $q$-profiles obtained by \textsc{cronos} interpretative simulations and the measured $q$-profiles is satisfactory, as seen in Fig.~\ref{fig:figure2}. The average discrepancy between the interpretative and measured $q$-profile values at $\rho=0.33$ and $0.64$ is $\sim10\%$, within the estimated uncertainty of the $q$-profile measurements. The experimental $q$-profiles were obtained by EFIT constrained by either Faraday rotation measurements (discharges 70084 and 73221) or motional Stark effect (MSE) measurements (discharges 66130 and 66404). 

In the \textsc{Gene} simulations, typical grid parameters were as follows: perpendicular box sizes $[L_x,L_y]=[170,125]$ in units of $\rho_s{\equiv}c_s/\Omega_{ci}=\sqrt{T_e/m_i}/\left(eB/m_i\right)$, perpendicular grid discretisations $[n_x,n_y]=[192,48]$, 24 point discretisation in the parallel direction, 32 points in the parallel velocity direction, and 8 magnetic moments. Extensive convergence tests were carried out for representative simulations throughout the parameter space spanned in this work. The lack of convergence of the heat fluxes with increasing $n_y$ as reported for \textsc{gyro}~\cite{cand03} simulations of discharge 70084 in Ref.\cite{mant11} - associated with increasing zonal flows - was not encountered here. In our cases the convergence with $n_y$ was well behaved. The difference may stem from the different treatment of the radial boundary conditions in the \textsc{Gene} and \textsc{gyro} simulations. Further investigation is necessary to ascertain this. The heat fluxes shown in the following sections are in gyroBohm normalised units, $q_{GB}=T_i^{2.5}n_im_i^{0.5}/e^2B^2R^2$. $k_y$ is in units of $1/\rho_s$. These heat fluxes correspond to time averaged values over the saturated state of the \textsc{Gene} simulations. The statistical flux variations due to intermittency are for clarity not explicitly shown as error bars. This variation is typically $5-10\%$ for our parameters. $\gamma$ and $\gamma_E$ are in units of $c_s/R$ where $c_s\equiv\sqrt{T_e/m_i}$. All rotation is considered to be purely toroidal unless specifically mentioned otherwise. For the low and high rotation discharges $\gamma_E=0.1$ and $0.3$ respectively, at both $\rho=0.33$ and $\rho=0.64$. These are representative $\gamma_E$ values for the low and high stiffness discharges from the dataset in Ref.\cite{mant11}.

%% file: sect3.tex
\section{Stiffness study at inner radius $\rho=0.33$}
\label{sec:lowshear}
In this section, we isolate the effect of various parameters on ion profile stiffness and critical threshold, at $\rho=0.33$ (where the transition to low stiffness at high rotation was observed). These parameters are: $q$, $\hat{s}$, rotation, effect of rotation on the magnetohydrodynamic (MHD) equilibrium, fast particle content, $R/L_n$, $\beta_e$, and $Z_{eff}$. %We then proceed to realistic simulations of 70084, 66130, 66404 and 73221 for a full comparison between the gyrokinetic predictions and experimental ion heat fluxes. These simulations simultaneously include: numerical geometry, collisions, electromagnetic effects, $Z_{eff}$ (by including a carbon species), and realistic $T_e/T_i$. 

%We however begin by first looking at a similar case to the GYRO s=0.6 run for 70084 as seen in Fig.~8c in Ref.\cite{mant11}. This is seen in Fig.~\ref{fig:sec3_s06}. The simulations are electrostatic, collisionless, with s/q=0.6/1.3, and with circular geometry. Two cases were simulated, with $\gamma_E=0,0.3$. Note the different definition of $\gamma_E$ here compared with in Ref.\cite{mant11}. Here we normalize by the major radius instead of the minor radius, introduced a difference of $\sim3$ in the quoted values of $\gamma_E$. 

%\begin{figure}[htbp]
	%\centering
		%\includegraphics[scale=0.6]{sec3_s06.eps}
		%\caption{Non-linear GENE $R/L_{Ti}$ scans for various levels of $\hat{s}$ and q-profile with circular geometry at x=0.33.}
	%\label{fig:sec3_s06}
%\end{figure}

\subsection{Stiffness and threshold sensitivity to $q$ and $\hat{s}$}
\label{sec:stiffsens}
\begin{figure}[htbp]
	\centering
		\includegraphics[scale=0.7]{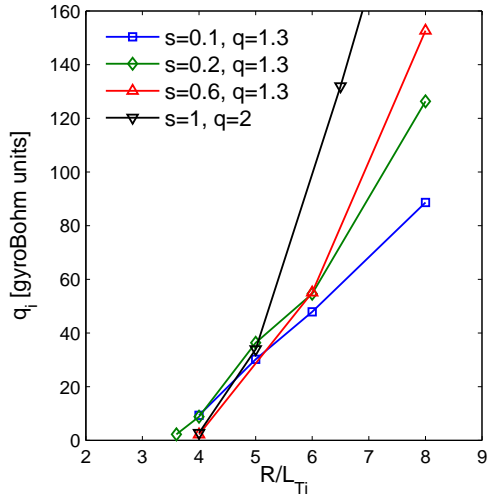}
		\caption{Non-linear electrostatic collisionless \textsc{Gene} $R/L_{Ti}$ scans for various levels of $\hat{s}$ and $q$-profile with circular geometry at $\rho=0.33$. $R/L_{Te}=5$, and $R/L_{n}=1.1$.}
	\label{fig:figure3}
\end{figure}

While the linear ITG turbulence threshold increases with $\hat{s}/q$~\cite{jenk01}, the stiffness (i.e. the rate of change of the gyro-Bohm normalised ion heat flux with respect to $R/L_{Ti}$) decreases in non-linear ITG simulations with both decreasing $\hat{s}$ (for $\hat{s}<\sim0.7$) and decreasing $q$. The reduced stiffness for decreasing $\hat{s}$ at low-$\hat{s}$ has been shown to be correlated with increased coupling to zonal flows~\cite{citr12}. For decreasing $q$, the stiffness reduction is due to a decreased downshift (compared with the peak in the linear spectrum) in the peak wavenumber of the turbulence spectrum, indicating decreased correlation lengths~\cite{dann05,hiro05,kins06}. These sensitivities are shown in Fig.~\ref{fig:figure3}. The stiffness level is shown to decrease for decreasing $\hat{s}$ at low-$\hat{s}$ at constant $q$=1.3. We can also see that for both the $\hat{s}/q=0.6/1.3$ and $\hat{s}/q=1/2$ cases the turbulence threshold is similar while the stiffness is lower for the $\hat{s}/q=0.6/1.3$ case, due to the decreased $q$.

We will deliberately make an optimistic assumption that $\hat{s}/q=0.2/1.3$ throughout all the subsequent parameter scans carried out at $\rho=0.33$ in this section. For the numerical geometry cases, this was done by modifying the current profile input into \textsc{cronos} such that at $\rho=0.33$ values of $\hat{s}/q=0.2/1.3$ were obtained following the solution of the Grad-Shafranov equation. The choice of assuming $\hat{s}/q=0.2/1.3$ is to ensure that we are in a `low-$\hat{s}$ regime', which has been hypothesised to be an important factor in the stiffness reduction, based on the observed correlation between low stiffness and low-$\hat{s}$ throughout the data set in Ref.\cite{mant11}. %An added advantage is that this choice is optimistic regarding stiffness reduction. Thus, if the non-linear gyrokinetic simulations do not display the same degree of low stiffness as experimentally observed even with $\hat{s}/q=0.2/1.3$ - as will indeed be shown - then this choice ensures that uncertainties in the $q$-profile cannot be invoked to suggest that the $\hat{s}$ and $q$ values used were not low \textit{enough} to obtain reduced stiffness. 

\begin{figure}[htbp]
	\centering
		\includegraphics[scale=0.7]{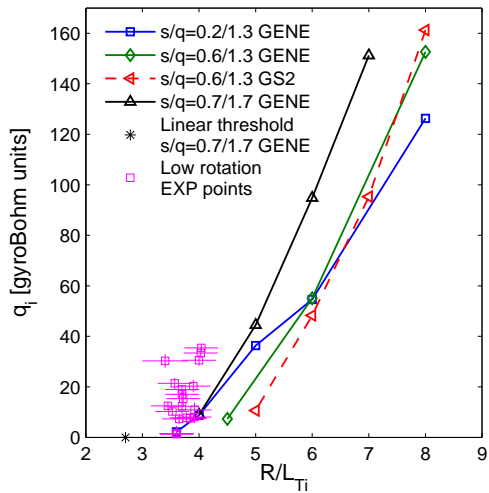}
		\caption{Comparison between non-linear electrostatic collisionless \textsc{Gene} and \textsc{gs2} $R/L_{Ti}$ scans with the low rotation data from the data-set in Ref.\cite{mant11} for various levels of $\hat{s}$ and $q$. The \textsc{Gene} runs are with circular geometry at $\rho=0.33$, the \textsc{gs2} runs with $\hat{s}-\alpha$ geometry. $R/L_{Te}=5$, and $R/L_{n}=1.1$.}
	\label{fig:figure4}
\end{figure}

The discussion of the sensitivity of the linear threshold to $q$ brings us to an important point. In Refs.\cite{mant09,mant11}, it was pointed out that the measured turbulence threshold of the low-rotation discharges in the data set were lower than the predicted non-linearly upshifted (Dimits shift)~\cite{dimi00} thresholds. These thresholds were predicted by non-linear simulations based on the low-rotation discharge 70084 performed with the \textsc{gs2} non-linear gyrokinetic code~\cite{kots95}. The measured turbulence thresholds agreed with the simulated linear thresholds as opposed to the non-linear thresholds. This result thus questioned the Dimits shift paradigm. The $q$ value used for these previous simulations was $q=1.3$, based on the processed data at the time. However, the data processing methodology for obtaining $q$-profiles using Faraday rotation constraints at JET~\cite{brix08} has since been improved, leading to a revision of the measured $q$-profile value to $q=1.7$ at $\rho=0.33$ for $t\sim10$~s for discharge 70084. The impact of this difference in $q$ on the linear and non-linear thresholds as predicted by the gyrokinetic codes is significant. This is shown in Fig.~\ref{fig:figure4}. In the figure, the \textsc{gs2} predicted ion heat fluxes for the $\hat{s}/q=0.6/1.3$ case (as shown in Ref.\cite{mant09}) is compared with the analogous \textsc{Gene} simulations. The agreement between the codes is good, apart from the zone near the threshold. This difference is likely to be due to the different methods used to calculate the geometry: analytical circular in \textsc{Gene}, and $\hat{s}-\alpha$ geometry in \textsc{gs2}. However, the non-linear threshold for $\hat{s}/q=0.6/1.3$ in both codes is approximately $R/L_{Ti}\sim4.5$, above the experimental threshold from Ref.\cite{mant11}. These curves can then be compared with the $R/L_{Ti}$ scan (carried out with \textsc{Gene}) with the revised, lower turbulence threshold corresponding to $\hat{s}/q=0.7/1.7$. In this case, the linear threshold is $R/L_{Ti}=2.7$, and the non-linear threshold following the Dimits shift is at $R/L_{Ti}\sim3.5-4$, in much better agreement with the experimental data. Consistency of the $\hat{s}/q=0.7/1.7$ values with both the revised experimental $q$-profile and \textsc{cronos} simulations is thus suggestive that the Dimits shift paradigm is in fact now supported by the experimental observations. However, the high sensitivity of the turbulence thresholds to the precise $\hat{s}$ and $q$ values leads us to a more conservative conclusion that no firm statement is justified regarding the consistency of the experimental data with the non-linear Dimits shift. The various values of $\hat{s}$ and $q$ used in the $R/L_{Ti}$ scans in Fig.~\ref{fig:figure4} (including the $\hat{s}/q=0.2/1.3$ values subsequently used in this section) can be seen to constitute a sensitivity test of the `reasonable' range of $q$ and $\hat{s}$ in lieu of rigorous error bars. The one clear conclusion from this sensitivity scan, is that there is no clear \textit{disagreement} between the experimental data and the non-linear threshold upshifted due to the Dimits shift.

We note that the observed Dimits shift for the $\hat{s}/q=0.7/1.7$ case is $\frac{{\Delta}\left(R/L_{Ti}\right)}{R/L_{Tcrit}}\approx25\%$. This value is comparable with the $\frac{{\Delta}\left(R/L_{Ti}\right)}{R/L_{Tcrit}}\approx20\%$ shifts observed in previous realistic simulations with kinetic electrons~\cite{mikk08}. These shifts are significantly reduced compared with adiabatic electron simulations, where shifts of up to $50\%$ are observed. The relatively low magnitude of the Dimits shift in simulations with realistic parameters illustrates that a definitive experimental observation of the effect may be extremely challenging, due to the error bars associated with the extrapolation to a critical turbulence threshold.

While the non-linear turbulence threshold extrapolated from the $\hat{s}/q=0.7/1.7$ curve in Fig.~\ref{fig:figure4} matches the experimental threshold, the simulated stiffness level is seemingly lower than the experimental trend. The possibility that this discrepancy can be explained by the differences in $T_e/T_i$ between the low flux and high flux points in the low rotation branch - which impact the critical threshold - is explored in the more comprehensive simulations shown in section~\ref{subsubsec:highstiff}. %This is the essence of question (4) outlined in section~\ref{sec:intro}.

\subsection{Stiffness sensitivity to rotation}
\begin{figure}[htbp]
	\centering
		\includegraphics[scale=0.7]{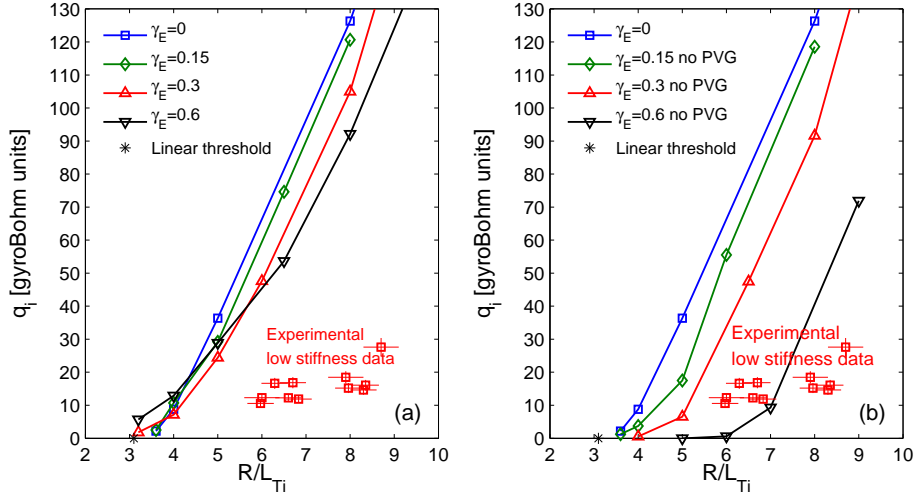}
		\caption{Non-linear \textsc{Gene} $R/L_{Ti}$ scans based on 70084 parameters at $\rho=0.33$ ($q/\epsilon=11.8$ for circular geometry) and various levels of $\gamma_E~[c_s/R]$. Runs including PVG destabilisation are shown in (a). Runs ignoring PVG destabilisation are seen in (b). All runs were electrostatic, collisionless, and with circular geometry.  The results are compared with the low stiffness data at $\rho=0.33$ from Ref.\cite{mant11}.}
	\label{fig:figure5}
\end{figure}

In this subsection we isolate the effect of rotation on stiffness, assuming pure toroidal rotation. This assumption is justified for JET discharges with significant NBI. Collisionless, electrostatic simulations based on 70084 parameters (assuming $\hat{s}/q=0.2/1.3$) are carried out, applying analytical circular geometry~\cite{lapi09}. The predicted gyroBohm normalised ion heat fluxes from the $R/L_{Ti}$ scans are shown in Fig.~\ref{fig:figure5}. The sensitivity to $\gamma_E$ is examined when including (Fig.~\ref{fig:figure5}a) and neglecting (Fig.~\ref{fig:figure5}b) the contribution from parallel velocity gradient (PVG) modes. Even for $\gamma_E=0.6$, double the highest level of flow shear achieved in the reference data set from Ref.\cite{mant11}, the simulated level of reduced stiffness is significantly less than the experimental observation, as seen by the direct comparison with the reference data. However, interesting effects related to the competition between stabilising $E{\times}B$ shear and destabilising PVG modes - particularly in the vicinity of the threshold - are observed. At low $R/L_{Ti}$, the PVG destabilisation can dominate over the ITG turbulence, reducing stiffness in that region of parameter space. Due to the PVG destabilisation, the fluxes do not continue to decrease towards the ITG instability thresholds. This is seen in Fig.~\ref{fig:figure5}a by examining the various curves at fixed $R/L_{Ti}$. At low $R/L_{Ti}$, the fluxes rise with $\gamma_E$ due to PVG drive. However at higher $R/L_{Ti}$, the fluxes decrease with $R/L_{Ti}$ due to the ITG stabilisation by perpendicular $E{\times}B$ flow shear dominating over the PVG destabilisation. In Fig.~\ref{fig:figure5}b the parallel velocity gradients were artificially removed from the system, and the picture reverts to a threshold shift. Note that particularly for the (red) $\gamma_E=0.3$ and (black) $\gamma_E=0.6$ curves the apparent reduced slope near threshold is not necessarily indicative of reduced stiffness in that regime, since the actual effective non-linear threshold may lie between the precise values of the $R/L_{Ti}$ values chosen for the simulations.

\begin{figure}[htbp]
	\centering
		\includegraphics[scale=0.6]{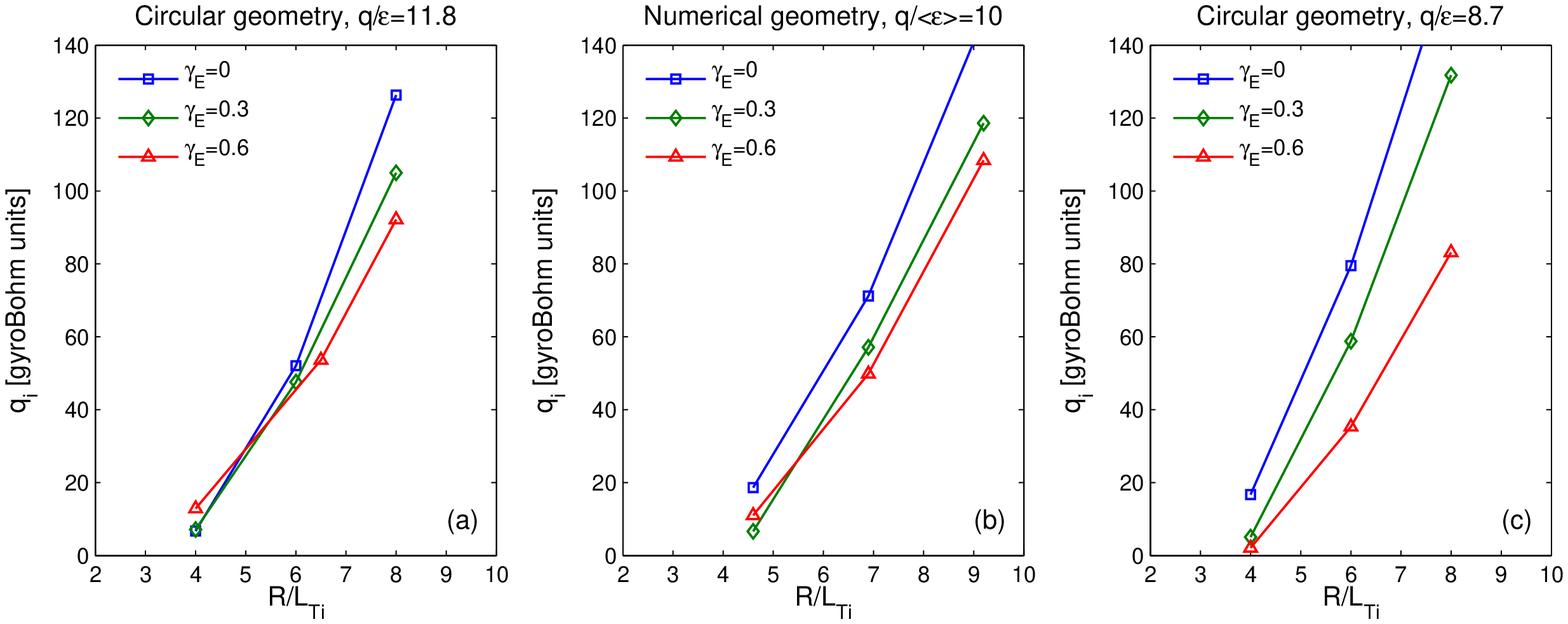}
		\caption{$q/\epsilon$ sensitivity of the PVG destabilisation as seen in $R/L_{Ti}$ scans of ion heat flux. As $q/\epsilon$ is progressively raised, the $\gamma_E$ induced stabilisation can not only be reduced but can even be reversed in the region of the instability threshold. Runs were electrostatic, collisionless, and with circular geometry.}
	\label{fig:figure6}
\end{figure}

For pure toroidal rotation, the relative importance of PVG destabilisation versus $E{\times}B$ stabilisation is sensitive to the geometric parameter $q/\epsilon$ (where $\epsilon{\equiv}r/R$)~\cite{high10}. As $q/\epsilon$ increases, the field lines are increasingly projected onto the toroidal direction. In Fig.~\ref{fig:figure6}, a $q/\epsilon$ scan is carried out by varying $\epsilon$ in the various $R/L_{Ti}$ scans. Simulations with $\epsilon=0.11,0.15$ assuming circular geometry were performed, as well as an $\langle\epsilon\rangle{\equiv}{\langle}r{\rangle}/R=0.13$ case from the flux surface averaged minor radius at $\rho=0.33$ using numerical geometry from the \textsc{helena}~\cite{huys91} equilibrium in the \textsc{cronos} simulation of discharge 70084. The $R/L_{Ti}$ values in the plots corresponding to numerical geometry are defined here with respect to the averaged midplane minor radius. The relative strength of the PVG destabilisation is seen to weaken as expected with decreasing $q/\epsilon$, until an almost pure threshold shift case is reached with $q/\epsilon=8.7$. %While this qualitative PVG dependence on $q/\epsilon$ is well known, we here make quantitative predictions with kinetic electrons for $R/L_{Ti}$ values in the vicinity of the turbulence threshold. For our particular case, the 70084 $\rho=0.33$ parameters are very close to the boundary of a `zero-turbulence-manifold' as defined in Ref.\cite{high12}. That study was carried out with adiabatic electrons, which may decrease the predicted $\gamma_E$ necessary for turbulence quenching compared with the kinetic electron case due to the reduced trapped electron drive.  

%The potential importance of the PVG modes in the high $q/\epsilon$ regime is also seen in the linear results shown in Fig.~\ref{fig:sec3_rotlin}. Low and high s/q cases with and without the inclusion of PVG modes are shown. In the q=2 case, the growth rates at high $\gamma_E$ even exceed the $\gamma_E=0$ growth rate, while for the q=1.3 case (more realistic at $\epsilon=0.11$, the inclusion of PVG significantly reduces the degree of mode suppression brought about by the rotation.

The interplay between PVG destabilisation and $E{\times}B$ stabilisation demands that PVG modes are correctly accounted for in reduced transport models - such as in gyrokinetic or gyrofluid based quasilinear models. Correct modelling near the turbulent thresholds is particularly critical for high temperature tokamaks, such as ITER. This is because the normalised fluxes are expected to be in the vicinity of the turbulence thresholds due to the $T_i^{5/2}$ normalisation dependence. 

%\begin{figure}[htbp]
%	\centering
%		\includegraphics[scale=0.7]{sec3_rotlin.eps}
%		\caption{Linear GENE $\gamma_E$ scans of 70084 with $\epsilon=0.11$. Cases with s/q=0.2/1.3 (red curves) and s/q=1/2 (blue curves) with (solid %curves) and without (dashed curves) the inclusion of PVG modes are shown.}
%	\label{fig:sec3_rotlin}
%\end{figure}

Finally, we note that the observed Dimits shift for these cases is only ${\Delta}\left(R/L_{Ti}\right)\approx0.5$, or alternatively $\frac{{\Delta}\left(R/L_{Ti}\right)}{R/L_{Tcrit}}\approx15\%$. This is another example of the relatively small Dimits shift observed in realistic simulations with kinetic electrons, as also shown in section \ref{sec:stiffsens} and in Ref.~\cite{mikk08}. The linear threshold shown in Fig.~\ref{fig:figure5} was calculated by extrapolation to zero-growth-rate of linear $R/L_{Ti}$ scans with \textsc{Gene}. The linear threshold in the numerical geometry case is nearly identical to the circular geometry case. 

In summary, the \textsc{Gene} simulations do not predict a significant reduction in stiffness due to flow shear, even with our deliberate choice of $\hat{s}=0.2$. As suggested by Fig.~\ref{fig:figure6}c and as shown in section~\ref{subsec:all}, a significant reduction of flux due to flow shear is only seen when both the effect of PVG destabilisation is artificially reduced, and $\gamma_E$ is increased beyond the experimental values expected from the toroidal flow shear.

\subsection{Effect of rotation on the equilibrium}
\begin{figure}[htbp]
	\centering
		\includegraphics[scale=0.7]{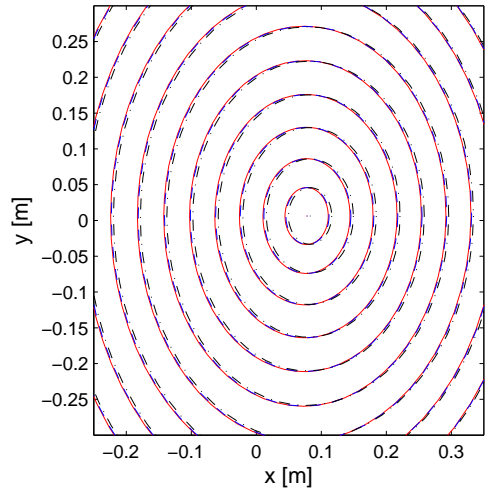}
		\caption{Flux surfaces in the vicinity of the magnetic axis from a solution of the generalised Grad-Shafranov equation using the kinetic profiles of 70084 and scaled rotation profiles from 66404. Three cases are shown: static (red solid curves), $\gamma_E=0.3$ (blue dashed curves) and $\gamma_E=0.6$ (black dashed-dotted curves).}
	\label{fig:figure7}
\end{figure}

In the previous section we examined the direct impact of rotation on the ion-heat-flux level through the flow shear. In this section we examine an indirect effect of rotation on the turbulent system through the impact of the centrifugal force on the plasma equilibrium. An extended Grad-Shafranov equation including toroidal rotation was solved with the \textsc{finesse} code~\cite{beli02}, using the 70084 pressure and F profiles as input, where $F{\equiv}B_{tor}R$. For the rotation profiles, scaled variants from 66404 were used such that static ($\gamma_E=0$), $\gamma_E=0.3$ and $\gamma_E=0.6$ cases were studied. All values correspond to $\rho=0.33$. The different equilibria are seen in Fig.~\ref{fig:figure7}. The sensitivity of the equilibria to these levels of rotation are found to be small, as expected due to the Mach number squared scaling of the `rotation pressure'. Only a 10\% increase in the Shafranov shift was observed between the static and $\gamma_E=0.6$ case. The non-linear predicted flux sensitivity to this different Shafranov shift is also minimal, with only a 6\% decrease in ion heat flux when the $\gamma_E=0.3$ equilibrium is used compared with the static equilibrium for a run with $R/L_{Ti}=6.9$. We can thus conclude that the effect of rotation on the equilibrium itself can only play a minor role in setting the profile stiffness.

\subsection{Inclusion of fast particles}
\label{sec:fast}
The discharges studied are relatively low density cases. This may allow for the sustainment of a significant fraction of non-thermalised fast ions in the plasma, particularly for the higher rotation cases, where significant NBI is employed. The impact of these fast ions on the ion-heat-flux is investigated in this section. The presence of fast ions is in general predicted to reduce the turbulent drive through a number of mechanisms. One such mechanism is the dilution of the main ion species by the fast ions. In ASDEX Upgrade strong evidence has pointed to a fast ion dilution mechanism for ITB formation at low density~\cite{tard08}. In addition, an increase in local $\alpha=\beta'q^2R$ due to fast ion suprathermal pressure gradients also stabilises ITG modes through electromagnetic effects. This has been suggested as a mechanism for ITB formation in low density JET hybrid discharges~\cite{roma10}. Finally, particularly at low magnetic shear, a geometric stabilization mechanism exists whereby the increased Shafranov shift induced by increased $\alpha$ modifies the drift frequencies and reduces the drive of ITG instabilities~\cite{bour05}. A fast ion fraction has been previously proposed to be responsible for mismatch between gyrokinetic simulations and experiments~\cite{holl11}.

Monte Carlo simulations of the NBI injection and subsequent fast ion slowing down were carried out for discharge 66404 with \textsc{nemo/spot}~\cite{schn11} within the \textsc{cronos} integrated modelling framework. An average fast particle energy ($\approx35keV$) at $\rho=0.33$ was calculated. In the \textsc{Gene} simulations, the fast particle temperature was approximated to the average fast particle energy value. Approximating the fast particle slowing-down distribution as a Maxwellian is not strictly justified. However, the high energies (compared to the main ions) of the fast particles leads to a significant proportion of the fast particles having Larmor radii greater than the typical turbulent eddy scale lengths. This then decreases the backreaction of the fast tail on the system. However, a dedicated study of the impact of various fast particle distribution functions on the turbulent system is necessary to fully justify this assumption.

\begin{figure}[htbp]
	\centering
		\includegraphics[scale=0.7]{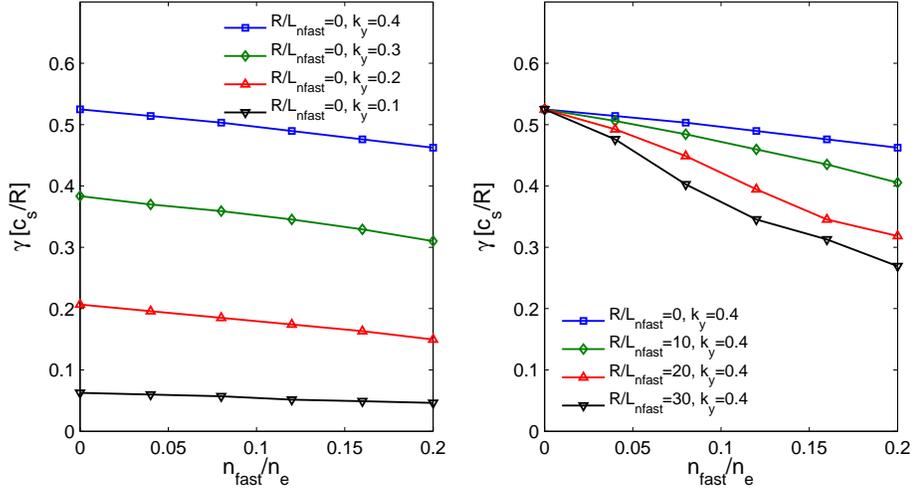}
		\caption{Linear fast particle density scans with 66404 parameters at $\rho=0.33$ at various values of $k_y$ (left panel) and $R/L_{nfast}$ (right panel). Runs were electromagnetic, with collisions, and with numerical geometry.}
	\label{fig:figure8}
\end{figure}

A linear \textsc{Gene} scan of fast particle densities (relative to $n_e$) can be seen in Fig.~\ref{fig:figure8}. The scan is carried out for various $k_y$ values in Fig.~\ref{fig:figure8}a. The scans assume $R/L_{Tfast}=R/_{nfast}=0$ - equivalent to assuming pure ion dilution. The $R/L_{nfast}$ sensitivity is examined in Fig.~\ref{fig:figure8}b at $k_y=0.4$. Increasing $R/L_{nfast}$ corresponds to an increased pressure gradient, increasing the stabilisation through electromagnetic effects as expected. The modelled fast ion pressure gradient at $\rho=0.33$ corresponds to $R/L_{nfast}=15$. A suppression of the growth rates is observed with increasing $n_{fast}/n_e$. However, for discharge 66404 (high rotation, low stiffness case), the fast ion fraction is predicted by \textsc{nemo/spot} to be only $\sim10\%$. Interpolating to $R/L_{nfast}=15$, this corresponds to a growth rate reduction of $\sim15\%$. According to the linear simulations, this magnitude is insufficient to explain the reduced stiffness. 

The above analysis was carried out for the NBI fast ions. With regard to the ICRH fast ions, the effect of the ion dilution is indirectly taken into account in the full modelling described in section \ref{subsec:all} by the higher $Z_{eff}$ due to the $He3$ minority. From SELFO~\cite{hedi02} modelling, which includes finite ion cyclotron orbit width effects, important for an accurate calculation of the ICRH fast ion pressure profile width, we determined that the ICRH induced suprathermal pressure gradient is similar in magnitude to the NBI profile at $\rho=0.33$ with a similar linear stabilization effect. However, we anticipate the results of section~\ref{subsec:beta} and section~\ref{subsec:all} and state that the electromagnetic stabilization effect is enhanced non-linearly, and is a key factor in explaining the observed stiffness reduction.

%While the $n_{fast}/n_e$ ratio and pressure gradients may be low to explain the stiffness reduction, the absolute fast particle pressure due to NBI and ICRH can still be a significant fraction of the total pressure. For the NBI driven fast ions in discharge 66404 at $\rho=0.33$, $P_{fast}/P_{tot}=0.21$ according to \textsc{nemo/spot} modelling (averaging between 6.5-7.5~s).  

\begin{figure}[htbp]
	\centering
		\includegraphics[scale=0.7]{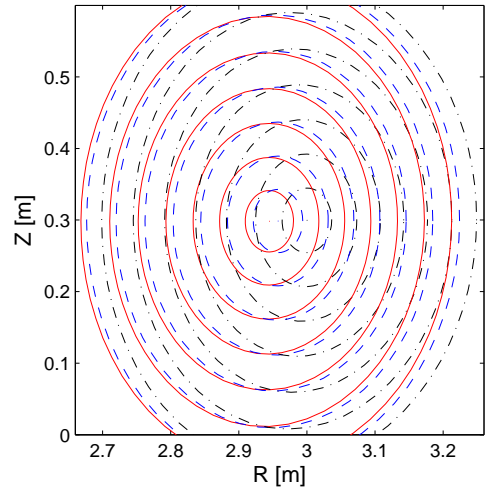}
		\caption{Flux surfaces in the vicinity of the magnetic axis for discharge 70084 (red solid curves), 66404 without fast particle pressure, (blue dashed curves) and 66404 with the inclusion of fast particle pressure (black dashed-dotted curves). $x=y=0$ corresponds to the geometric axis.}
	\label{fig:figure9}
\end{figure}

\begin{figure}[htbp]
	\centering
		\includegraphics[scale=0.7]{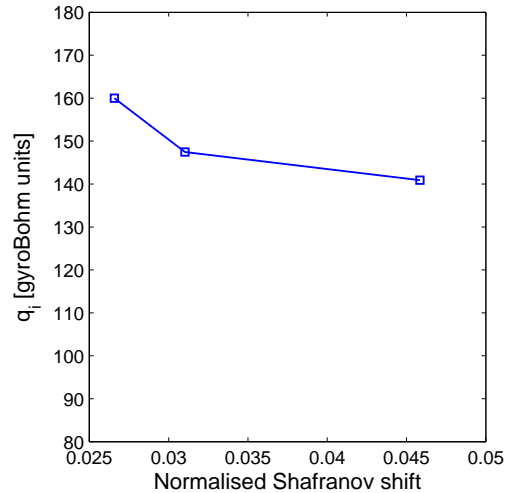}
		\caption{Flux reduction as a function of Shafranov shift normalised to the major radius for the three equilibria presented in Fig.~\ref{fig:figure9}. }
	\label{fig:figure10}
\end{figure}

In the above analysis, the influence of the suprathermal pressure on the magnetic geometry through an increased Shafranov shift was not taken into account. The increased Shafranov shift can be seen in Fig.~\ref{fig:figure9}, where the flux surfaces for the low power discharge 70084, and the high power discharge 66404 (with and without the inclusion of the NBI fast particle pressure) are compared. The fast particle contribution to the 66404 Shafranov shift is significant. For 70084, the Shafranov shift is $\approx7.5$~cm. For 66404 with the thermal pressure contribution only, the Shafranov shift is $\approx8.8$~cm. For 66404 with the total pressure (including fast particles), the total Shafranov shift is $\approx13$~cm. The impact of this difference on the predicted fluxes was investigated through dedicated non-linear simulations. The impact was observed to be not negligible but also not a dominating factor. A flux reduction of $15\%$ was observed in the non-linear simulations with $R/L_{Ti}=8$ when substituting the numerical geometry from 70084 with that of 66404 (i.e. with the fast particle content), as seen in Fig.~\ref{fig:figure10}. We can thus conclude that the effect of fast particles as seen in linear simulations, through both the increased pressure gradient and an increased Shafranov shift, is significant but cannot be the sole explanation for the reduction in stiffness observed. 

\subsection{Impact of $R/L_n$ on the stiffness level}
\begin{figure}[htbp]
	\centering
		\includegraphics[scale=0.7]{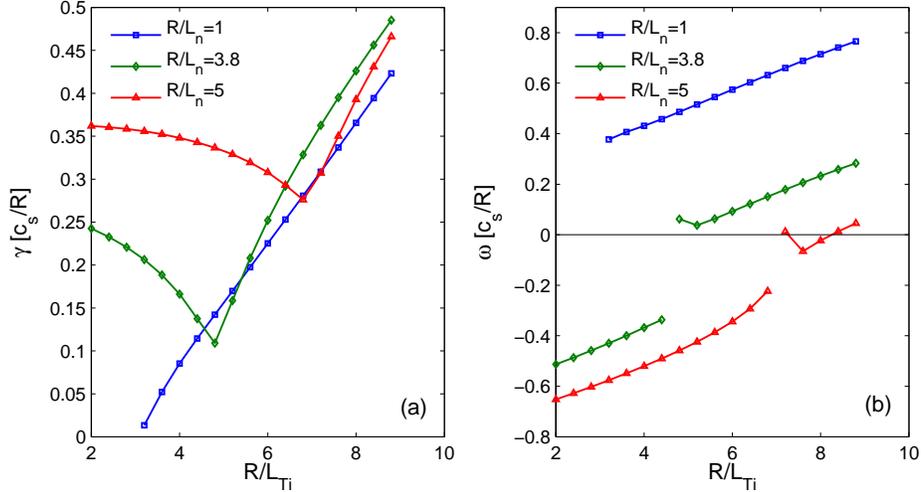}
		\caption{Linear $R/L_{Ti}$ scans based on 66404 parameters at $\rho=0.33$ with varying $R/L_n$. Growth rates are shown in (a), and frequencies in (b). Runs were electrostatic, with collisions, and with circular geometry.}
	\label{fig:figure11}
\end{figure}

In the limited experimental data set studied, there is a wide variation in $R/L_n$, from 1.4 in the 70084 case to 3.8 in the 66404 case (which corresponds to the highest $R/L_{Ti}$ in the data set). The sensitivity of the turbulence to the $R/L_n$ value was thus examined. In particular, the possibility that non-linear ITG-TEM (trapped electron mode) interplay takes place which can reduce the level of turbulence and thus the stiffness, as reported in Ref.\cite{merz10}, was investigated. In Fig.~\ref{fig:figure11}, these linear scans are shown. For $R/L_n=1$, the dominant mode propagates in the ion diamagnetic direction (ITG mode). However, for $R/L_n=3.8$ the mode at low $R/L_{Ti}$ propagates in the electron diamagnetic direction. This is most probably a density gradient driven TEM mode, which is stabilised by $R/L_{Ti}$ (which would correspond to low stiffness) until it switches to an ITG mode at $R/L_{Ti}\approx5$. At that point we would expect turbulence stabilisation according to Ref.\cite{merz10}. However, for higher $R/L_{Ti}$ the growth-rate stiffness is similar to the $R/L_n=1$ case, as a pure ITG regime is reached. For $R/L_n=5$ the TEM-dominated regime is maintained for much higher $R/L_{Ti}$. However, the highest experimental $R/L_n$ in the data set of Ref.\cite{mant11} is $R/L_n\approx4$. Furthermore, at the experimental high $R/L_{Ti}$ values the transport is ITG dominated and stiff even for $R/L_n=5$. Thus it is unlikely that $R/L_n$ is responsible for reduced profile stiffness. Furthermore, even if the stiffness is low, the actual growth rates themselves are high, and we may expect a high degree of transport. 

\begin{figure}[htbp]
	\centering
		\includegraphics[scale=0.7]{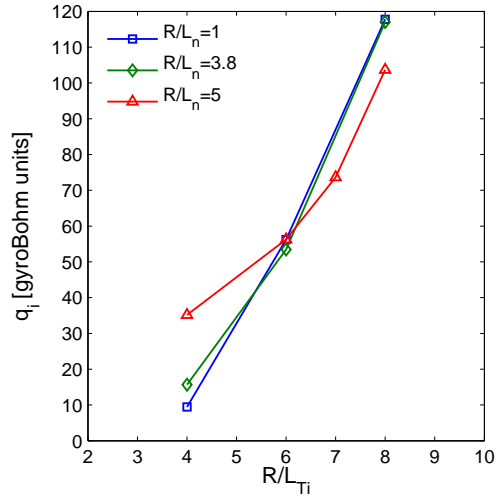}
		\caption{Non-linear $R/L_{Ti}$ scans based on 66404 parameters at $\rho=0.33$ with varying $R/L_n$. Runs were electrostatic, with collisions, and with circular geometry.}
	\label{fig:figure12}
\end{figure}

These results are maintained in the non-linear scans, seen in Fig.~\ref{fig:figure12}. While at lower $R/L_{Ti}$ stiffness is indeed reduced in the TEM regime for the high $R/L_n$ case, at higher $R/L_{Ti}$ values the difference in stiffness between the $R/L_n=1$ and $R/L_n=3.8$ cases becomes negligible. We can conclude that the variance of $R/L_n$ in the data set is not responsible for the observed difference in stiffness.

\subsection{Impact of $\beta_e$ on the stiffness level}
\label{subsec:beta}
In this subsection the sensitivity of the stiffness on electromagnetic effects - which arise for $\beta_e>0$ - is examined. The simulations carried out take discharge 66404 parameters as a reference. Linear (at $k_y=0.4$) and non-linear $\beta_e$ scans are shown in Fig.~\ref{fig:figure13}. From the linear scans, it is clear that the range of experimental $\beta_e$ values ($0-0.5\%$) are significantly below the kinetic ballooning mode (KBM) thresholds, characterised in the plot by the sharp upturn in growth rates at $\beta_e\approx1.5-2.4\%$; this finding is expected to carry over to the non-linear physics \cite{pues08}. Below the KBM threshold, $\beta_e$ stabilises the ITG mode~\cite{weil92}. For our parameters, this leads to a growth rate reduction of $\approx25\%$ at $\beta_e=0.5\%$. This is at the upper range of our experimental $\beta_e$ values. The $25\%$ growth rate stabilisation factor is not exceeded when repeating the linear simulations for $k_y=0.1-0.3$. The linear ITG mode is stabilised at lower and lower $\beta_e$ as $R/L_{Ti}$ is increased. This is likely due to the corresponding increase in $\alpha\equiv\beta'q^2R$, which can be considered a parameter of merit for the strength of the electromagnetic coupling. %For practical purposes, however, this should be of little benefit as the strong background gradient dependence of the $\beta_{KBM}$ threshold makes any ITG-stable regions irrelevant, since the KBM threshold significantly decreases with $R/L_{Ti}$.

A striking observation is that the \textit{non-linear} $\beta_e$ ITG stabilisation significantly exceeds the linear stabilisation. This is consistent with \textsc{Gene} results reported in Refs.~\cite{pues08,pues10}, as well as, to some degree, with other codes~\cite{chen03,cand05}. A decrease in ion heat flux by a factor of $65\%$ is seen in Fig.~\ref{fig:figure13}b for the $\gamma_E=0$, $R/L_{Ti}=9.2$ case between $\beta_e=0-0.48\%$. Simultaneously, while the ion heat flux is reduced by $\beta_e$ in the $\gamma_E=0$, $R/L_{Ti}=4.6$ case, it is not totally quenched. The observation that for $\beta_e>0$ the flux level is diminished over a range of $R/L_{Ti}$, yet is not totally quenched in the vicinity of the ITG threshold for $\beta_e=0$, is indicative that $\beta_e>0$ (within the range studied) induces a decrease in stiffness as opposed to a threshold shift. Note that the results reported in Ref.\cite{pues10} cannot be compared with those in Fig.~\ref{fig:figure13}b quantitatively, as TEM contributions to the overall turbulence picture may change in particular the $\beta_e$ dependence of the threshold shift.

It is interesting to note that the stabilising effect of flow shear is weakened by finite $\beta_e$ in the higher $R/L_{Ti}$ case, as seen in Fig.~\ref{fig:figure13}. In the $R/L_{Ti}=9.2$ case, the effect of flow shear on the turbulence switches from stabilising to destabilising as $\beta_e$ increases. However, in the $R/L_{Ti}=4.6$ case flow shear is always stabilising, and no discernible weakening of the stabilisation is seen as $\beta_e$ increases. Linearly, the PVG modes are not observed to lead to increased stabilisation at increased $\beta_e$. More effort needs to be taken in the future to uncover the non-linear effects which either increases PVG destabilisation or decreases the $E{\times}B$ stabilisation in the high $R/L_{Ti}$ case. %In general, any given value of $\gamma_E$ is expected to be relatively more stabilising for lower $R/L_{Ti}$ than higher $R/L_{Ti}$ in ITG turbulence, due to the lower growth rates and thus higher $\gamma_E/\gamma_k$ ratios (where $\gamma_k$ are the linear growth rates at each spatial scale), which is suggestive of an increase in the ratio between the flow shear decorrelation time and the non-linear autocorrelation time, leading to increased stabilisation. However, it is not clear why the stabilisation weakens at high $R/L_{Ti}$ with increasing $\beta_e$, and this is a topic that should be examined further.

We can thus conclude that electromagnetic effects play a significant role in stiffness reduction for our parameters, even at relatively low values of $\beta_e$. While this stiffness reduction is not sufficient to fully explain the experimentally observed stiffness reduction, it is a factor which must be taken into consideration. We note that the fast particle stabilisation observed in section \ref{sec:fast} is also an electromagnetic ITG stabilisation mechanism, and we can thus expect non-linearly a greater impact of the fast-particle stabilisation. This is explored in section \ref{subsec:all}.

\subsubsection{Non-linear electromagnetic stabilisation mechanisms}
From linear gyrokinetic analysis, electromagnetic stabilisation of ITG modes has been invoked as a possible factor in improved hybrid scenario confinement at ASDEX Upgrade and DIII-D, particularly at outer radii (i.e.~beyond half radius)~\cite{magg10}. The increased non-linear electromagnetic stabilisation reported here may point to an even greater importance of this effect than previously recognised. We note that the electromagnetic stabilisation is expected to be effective up to the recently discovered Non-Zonal Transition $\beta_e$ limit~\cite{pues13}, beyond which electromagnetic fluctuations effectively short out the zonal flows and lead to a significant increase in the saturated level of the ITG turbulent fluxes. This $\beta_e$ threshold very strongly depends on the background gradients, however, and for typical (low) gradients quickly becomes less restrictive than the KBM threshold. Coupled with the fact that this effect produces a limit with enormous stiffness, it can therefore be expected that standard experimental gradient and $\beta_e$ values in outer radii lie below this point, putting those cases in the electromagnetic stabilisation zone.

\begin{figure}[htbp]
	\centering
		\includegraphics[scale=0.7]{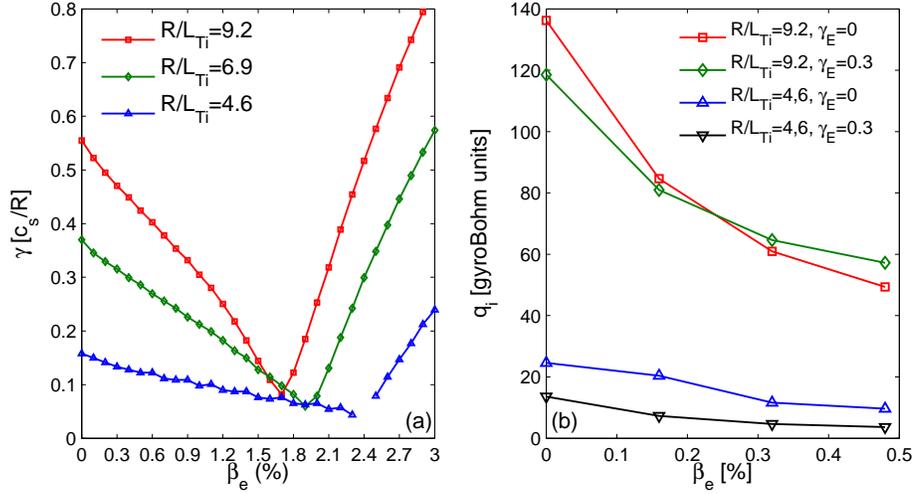}
		\caption{Linear (a) and non-linear (b) $\beta_e$ scans with 66404 parameters at $\rho=0.33$. $R/L_{Ti}$ and $\gamma_E$ are varied. Runs were with collisions and numerical geometry.}
	\label{fig:figure13}
\end{figure}

\begin{figure}[htbp]
	\centering
		\includegraphics[scale=0.7]{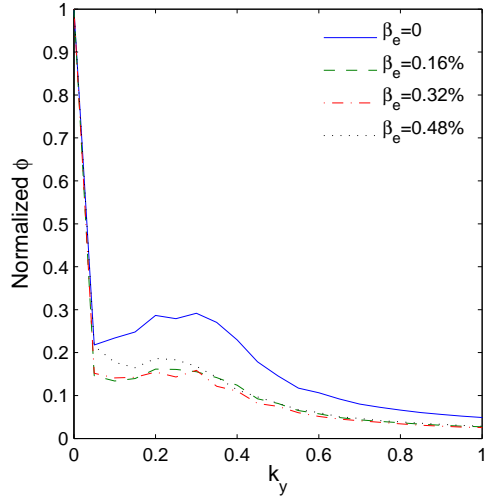}
		\caption{Amplitude spectra from the $\gamma_E=0$, $R/L_{Ti}=9.2$, non-linear $\beta_e$ scan displayed in Fig.~\ref{fig:figure13}b.}
	\label{fig:figure14}
\end{figure}

\begin{figure}[htbp]
	\centering
		\includegraphics[scale=0.7]{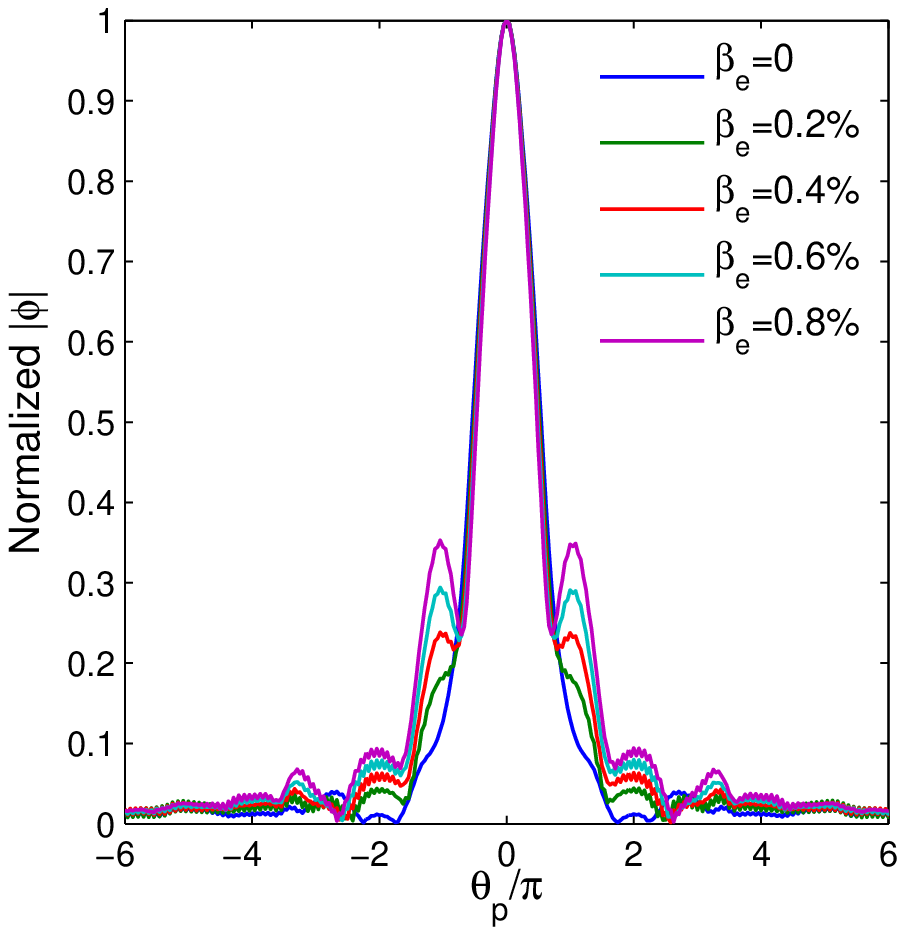}
		\caption{$\beta_e$ scan of ITG eigenmode structure calculated by linear-\textsc{Gene}. $R/L_{Ti}=9.2$, and $\gamma_E=0$.}
	\label{fig:figure15}
\end{figure}

In Refs.\cite{pues08,pues10}, an increase of the ratio between the zonal flow shearing rate to the unstable mode growth rate ($\omega_{ZF}/\gamma$) is observed with $\beta_e$. This is suggested to be part of the explanation of the non-linearly enhanced $\beta$-stabilisation. A possible physical mechanism for this relative increase in zonal flow activity, based on increased coupling between Alfv\'enic modes and drift waves, has also been suggested~\cite{mili11}. In Fig.~\ref{fig:figure14} we plot the mode amplitude spectra for the $\gamma_E=0$, $R/L_{Ti}=9.2$ scan over $\beta_e$ shown in Fig.~\ref{fig:figure13}. The amplitude spectra have been normalised to the zonal flow (or rather $k_y=0$, which constitutes a reasonably good measure) amplitudes. Indeed, a relative increase in the $k_y=0$ modes is seen for the electromagnetic cases, which may be related to the ITG $\beta_e$ stabilisation. Another possible mechanism for increased zonal flow coupling is the observed widening of the ITG linear eigenmode structure observed with increasing $\beta_e$, as shown in Fig.~\ref{fig:figure15}. The less ballooned structure facilitates the direct coupling to the poloidally symmetric zonal modes, similarly to what occurs at low magnetic shear~\cite{li04,citr12}. Further work is suggested to shed more light on this topic.

%We note that in Ref.\cite{pues10} a significant instability threshold shift was observed for $\beta_e>\sim0.4\%$. With our parameters, no such shift is seen across our spanned range of $\beta_e=0-0.48\%$. 

\begin{figure}[htbp]
	\centering
		\includegraphics[scale=0.7]{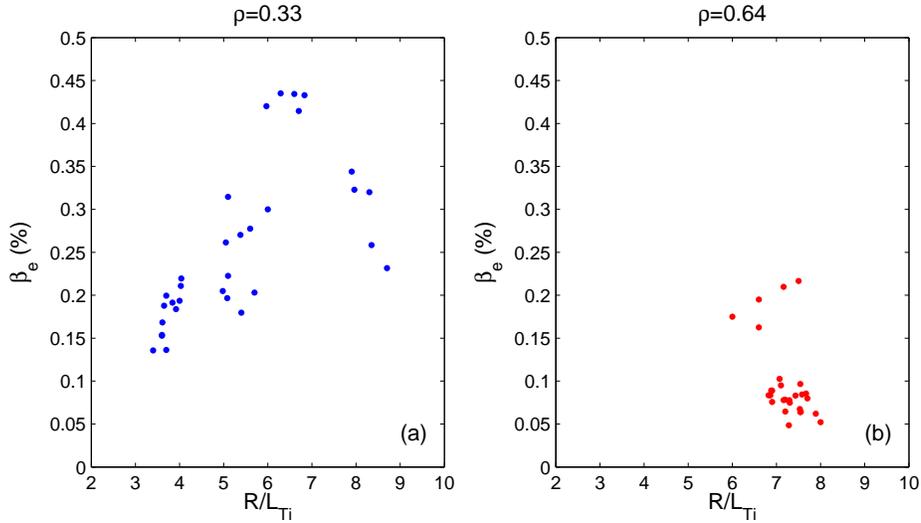}
		\caption{Correlation between $R/L_{Ti}$ and $\beta_e$ at $\rho=0.33$ (a) and at $\rho=0.64$ (b) from the entire data-set presented in Ref.\cite{mant11}.}
	\label{fig:figure16}
\end{figure}

In Fig.~\ref{fig:figure16} we can see, from the entire data set in Fig.~1 of Ref.\cite{mant11}, the correlations between $R/L_{Ti}$ and $\beta_e$ for $\rho=0.33$ and $\rho=0.64$. There is a generally limited but positive correlation between $\beta_e$ and $R/L_{Ti}$ at $\rho=0.33$, consistent with the reduced stiffness. At $\rho=0.64$, $\beta_e$ is generally much lower than at $\rho=0.33$. This is expected to further increase the stiffness at $\rho=0.64$ as observed experimentally, beyond the increase solely due to the higher $\hat{s}$ and $q$ values. We note that both at $\rho=0.33$ and $\rho=0.64$, a cluster of five points is visible at the highest respective $\beta_e$ values, separate from the main trend. These points correspond to discharges with significantly more heating power (between $10-15$~MW of NBI power) and slightly lower magnetic field (3~T as opposed to 3.4~T) than the rest of the dataset. The scatter that these points induce to the correlation between $\beta_e$ and $R/L_{Ti}$ is indicative of the difficulty in making a pure comparison of the effect of $\beta_e$ throughout the dataset, due to the concomitant changes in other plasma parameters and normalized ion heat fluxes.

%% file: sect4.tex
\section{Simulated and measured ion-heat-flux comparisons at $\rho=0.33$}
\label{subsec:all}
\begin{figure}[tbp]
	\centering
		\includegraphics[scale=0.7]{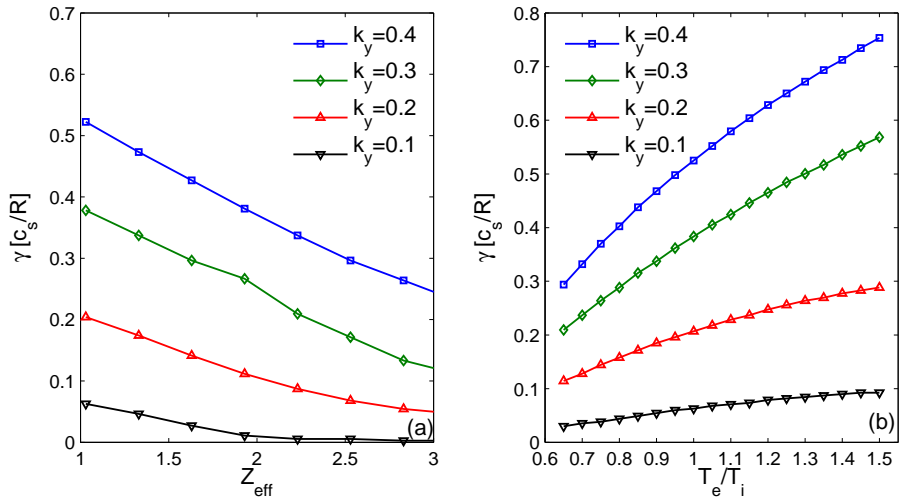}
		\caption{Sensitivity of growth rates to $Z_{eff}$ (a) and $T_e/T_i$ (b) from linear \textsc{Gene} runs based on 66404 parameters at $\rho=0.33$. Runs were electromagnetic, with collisions, and with numerical geometry.}
	\label{fig:figure17}
\end{figure}

In the previous section we have analysed the individual impact of numerous parameters on ITG mode stabilisation and ion temperature profile stiffness reduction. In this section we simultaneously include all effects, and carry out realistic simulations of all four discharges in the data set at $\rho=0.33$. We analyse the `high-stiffness-branch' and `low-stiffness-branch' separately in sections~\ref{subsubsec:highstiff} and \ref{subsubsec:lowstiff} respectively. Ion heat fluxes from non-linear simulations and experimental power balance are compared. The simulations included flow shear, the effect of rotation on equilibrium, experimental $R/L_n$, finite $\beta$, collisions, $Z_{eff}>1$, and experimental $T_e/T_i$. The effect of $Z_{eff}$ - which is stabilizing for ITG turbulence - was modelled in the 3-species simulations by lumping all impurities into a kinetic fully stripped carbon ion species. The carbon temperature, $R/L_{T}$, and $R/L_n$ were assumed the same as the main deuterium species. Simulations with varying $Z_{eff}$ values were carried out, to test the sensitivity of the predictions to the uncertainties in the $Z_{eff}$ profile shape. The growth rate sensitivity to $Z_{eff}$ and $T_e/T_i$ for linear \textsc{Gene} runs based on discharge 66404 can be seen in Fig.~\ref{fig:figure17}. For our cases, the $Z_{eff}$ stabilisation tends to be compensated by the $T_e/T_i>1$ destabilisation. In the non-linear simulations, assuming $R/L_{Tc}=0$ for the carbon species instead of $R/L_{Tc}=R/L_{Ti}$ altered the bulk ion heat flux by less than 2\%. %While the impact of the fast ions on the Shafranov shift is included in the simulations, fast ion dilution of the main ion species was not included. It was judged that the relatively minor ($\sim10\%$) impact of this effect on the simulated flux values did not justify the inclusion of the fast ions as a separate ion species in the simulations, which would significantly slow down the calculations.

\subsection{Investigation of the low-rotation, high-stiffness branch}
\label{subsubsec:highstiff}
In section~\ref{sec:stiffsens}, figure~\ref{fig:figure4}, it is evident that the stiffness of the simulated $\hat{s}/q=0.7/1.7$ curve is less than the apparent experimental trend. In this section, we examine the possibility that the higher $T_e/T_i$ of the high flux discharge 73221 in the low rotation branch is responsible for the increased flux, through the $T_e/T_i$ impact on the ITG critical threshold. It is important to note that this significant difference in $T_e/T_i$ between the high and low flux discharges in the low-rotation branch has become apparent only recently after data reprocessing following an in-vessel calibration of the ECE diagnostic. This is the reason why this aspect was not taken into account in Refs.~\cite{mant09,mant11}.

\begin{figure}[tbp]
	\centering
		\includegraphics[scale=0.7]{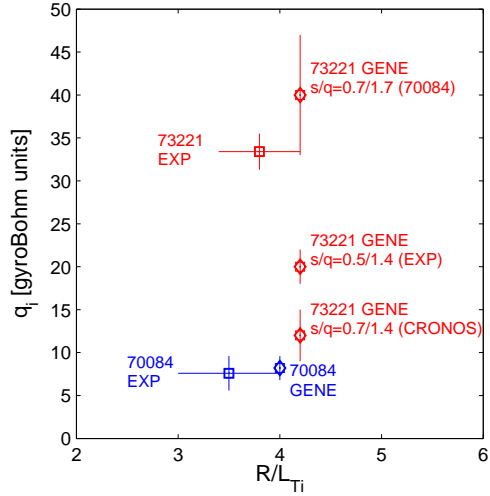}
		\caption{Comparison of experimental and simulated ion heat flux for discharges 70084 and 73221 situated on the `high-stiffness-branch' at $\rho=0.33$ from the dataset of Ref.\cite{mant11}. The 73221 simulation results shown were carried out for three separate $\hat{s}/q$ values to test the sensitivity to the $q$-profile uncertainties.}
	\label{fig:figure18}
\end{figure}

An $R/L_{Tcrit}\propto\left(1+T_i/T_e\right)$ scaling has been derived both analytically and from linear gyrokinetic simulations for the ITG instability~\cite{roma89,jenk01}. A decreased instability threshold leads to increased flux for a given $R/L_{Ti}$ value, as long as the stiffness level does not change with $T_e/T_i$. It has been predicted by non-linear simulations that the stiffness level is not highly sensitive to $T_e/T_i$ within the range relevant for our studied discharges~\cite{migl13}. The simulation results for the 70084 and 73221 discharges are shown in figure~\ref{fig:figure18}.  Since $R/L_{Ti}$ is close to threshold and the transport is relatively stiff, the results are highly sensitive to the input parameters. Additionally, the proximity to threshold leads to statistical flux variations due to intermittency often higher than the typical $5-10\%$ level observed for the simulations in this paper. These variations are displayed on the plot for these specific cases.  For 70084, agreement between the non-linear simulation and the experimental observation was found for reasonable departures from the base parameters recorded in table~\ref{tab:summary2}. $R/L_{Ti}$ and $T_e/T_i$ were both taken at the high end of their error bars. For the base values of $R/L_{Ti}$ and $T_e/T_i$, stability was predicted. $Z_{eff}$ was taken as 1.4, lower than $\langle{Z_{eff}}\rangle=2.2$. This is a reasonable assumption since the $Z_{eff}$ profiles tend to be hollow, and $\rho=0.33$ is relatively close to the magnetic axis. Making the same assumptions for 73221 (although maintaining the base value of $T_e/T_i$), the simulated flux value was found to be significantly lower than the experimental value. Even though $T_e/T_i$ is higher in 73221 than in 70084, the impact of the higher $T_e/T_i$ on the ITG critical threshold is compensated by the lower $q$ value calculated by the 73221 \textsc{cronos} interpretative simulation compared with 70084. However, when increasing the 73221 $q$ value in the simulation to equal the 70084 value - an increase of only $\sim15\%$ - the simulated flux value then becomes comparable to the experimental value. When assuming the Faraday rotation constrained EFIT $q$-profile for 73211, with $\hat{s}/q=0.5/1.4$, we obtain an intermediate flux level between the 70084 and 73221 experimental flux values. These tests of the variation in the 73221 flux values with variations of $q$ and $\hat{s}$ constitute a sensitivity analysis of the fluxes to reasonable estimates of the $q$-profile error bar. We thus deem that the $T_e/T_i$ increase of the high flux cases in this branch compared with the low flux cases is a likely explanation for the seeming anomalously high stiffness of this data-set. However, the high sensitivity of the simulated flux - through the impact on the critical threshold - to $T_e/T_i$ and the $q$-profile variations within the estimated experimental error bars precludes a firm conclusion on this point. The result lies within the uncertainties - particularly of the $q$-profile calculations. In table~\ref{tab:summary_sec3_1} we show the results for all simulations carried out for 70084 and 73221 - beyond those shown in Fig.~\ref{fig:figure18}. The sensitivities of the flux to input parameters such as $\hat{s}$, $q$, $Z_{eff}$, $\gamma_E$, and $R/L_n$ are shown. We note that the results are \textit{not} highly sensitive to wide variations in the $R/L_n$ values. 

\begin{table}
\small
\centering
\caption{\footnotesize Input data and ion heat flux results for discharge 70084 and 73221 non-linear simulations. The cases in bold font are the simulations displayed in Fig.~\ref{fig:figure18}.}
%\vspace{0.15cm}
\tabcolsep=0.11cm
\scalebox{0.7}{\begin{tabular}{c|c|c|c|c|c|c|c|c}
\label{tab:summary_sec3_1}
Shot number			  & $Z_{eff}$ & $R/L_{Ti}$ & 	$R/L_n$	 &   $T_e/T_i$ & $\gamma_E$ &  $\hat{s}$ & $q$ & $q_i$ [gyroBohm units] \\
\hline
70084 						&  1.4			& 3.5				 & 		1.4		 &		1.12     & 		0.07		&  		0.7    & 	1.7&		0					  \\
70084 						&  1.4			& 4 				 & 		1.4		 &		1.12     & 		0.1 		&  		0.2    & 	1.3&		0					  \\
\bf 70084 				& \bf 1.4		& \bf 4 		 & 	\bf	1.4	 &		\bf1.12  & 	\bf	0.1 	&  	\bf	0.7  & 	\bf1.7&	$\mathbf{8.2\pm1.4}$ \\
70084 						&  1.4			& 4 				 & 		1.4		 &		1.12     & 		0.07 		&  		0.7    & 	1.7&		$14\pm4$    \\
70084 						&  1.4			& 4 				 & 		1.4		 &		1.08     & 		0.1 		&  		0.7    & 	1.7&		0			      \\
70084 						&  1.9			& 4 				 & 		1.4		 &		1.12     & 		0.07 		&  		0.7    & 	1.7&		0			      \\
70084 						&  1.9			& 4 				 & 		1.4		 &		1.12     & 		0.04 		&  		0.7    & 	1.7&		$7.5\pm1.5$ \\
\hline
\hline
\bf 73221 				&  \bf1.4		& \bf4.2		 & 		\bf2.8 &		\bf1.35  & 		\bf0.02	&  	\bf	0.7  & 	\bf1.4 	&	 $\mathbf{12\pm3}$  \\
73221 						&  1.4			& 4.2 			 & 		1.0		 &		1.35     & 		0.02 		&  		0.7    & 	1.7		  &	 $48\pm2$ 		  		\\
\bf 73221 				&  \bf1.4		& \bf4.2		 & 	\bf	2.8	 &	\bf	1.35   & 	\bf	0.02	&  	\bf	0.7  & \bf	1.7 &	 $\mathbf{40\pm7}$  \\
73221 						&  1.4			& 4.2 			 & 		3.8		 &		1.35     & 		0.02 		&  		0.7    & 	1.7		  &	 $31\pm7$ 				  \\
\bf 73221 				&  \bf1.4		& \bf4.2		 & 	\bf	2.8	 &	\bf	1.35   & 	\bf	0.02	&  	\bf	0.5  & \bf	1.4 &	 $\mathbf{20\pm2}$  \\
73221 						&  1.9			& 4.2 			 & 		2.8		 &		1.35     & 		0.02 		&  		0.7    & 	1.4		  &	 $1.7\pm0.3$ 		    \\
73221 						&  1.9			& 4.2 			 & 		2.8		 &		1.35     & 		0.02 		&  		0.7    & 	1.7		  &	 $13\pm3$ 				  \\
\hline
\end{tabular}}
\end{table}

\subsection{Investigation of the high-rotation, low-stiffness branch}
\label{subsubsec:lowstiff}
\begin{figure}[tbp]
	\centering
		\includegraphics[scale=0.7]{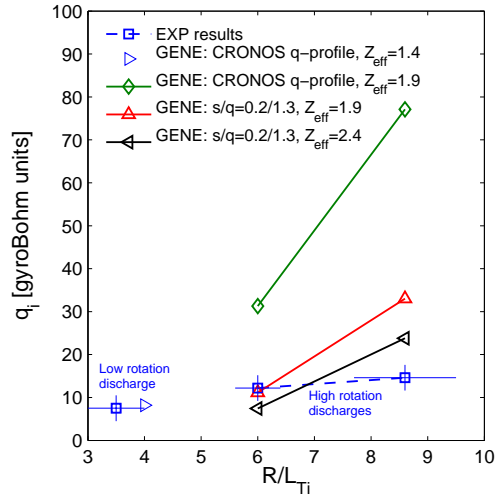}
		\caption{Comparison of non-linear simulations and experimental results for the three separate discharges at $\rho=0.33$. For the high rotation discharges, various sets of simulations with varying $\hat{s}$, $q$, and $Z_{eff}$ assumptions are shown. The \textsc{Gene} simulations corresponding to the low and high rotation discharges were carried out with $\gamma_E=0.1$ and $0.3$ respectively.}
	\label{fig:figure19}
\end{figure}

The comparison between the \textsc{Gene} non-linear simulations and the experimental heat fluxes for the `low-stiffness branch' is shown in Fig.~\ref{fig:figure19}. For the high rotation discharges, three separate sets of simulations are shown: with the nominal $q$-profile from the \textsc{cronos} interpretative runs and $Z_{eff}=1.9$, with the optimistic $\hat{s}/q=0.2/1.3$ assumption and $Z_{eff}=1.9$, and finally with the optimistic $\hat{s}/q=0.2/1.3$ assumption and $Z_{eff}=2.4$. Fast ions are not included in these simulations due to the added computational expense. The input parameters and flux values for these simulations, as well as additional simulations carried out for further sensitivity studies and for clarity not shown in Fig.~\ref{fig:figure19}, are listed in table~\ref{tab:summary_sec3_2}.

\begin{table}
\small
\centering
\caption{\footnotesize Input data and ion heat flux results for discharge 66130 and 66404 non-linear simulations. The cases in bold font are the simulations displayed in Fig.~\ref{fig:figure19}.}
%\vspace{0.15cm}
\tabcolsep=0.11cm
\scalebox{0.7}{\begin{tabular}{c|c|c|c|c|c|c}
\label{tab:summary_sec3_2}
Shot number			  & $Z_{eff}$ & $R/L_{Ti}$ & 	$T_e/T_i$ & $\hat{s}$ & $q$ & $q_i$ [gyroBohm units] 		 \\
\hline
66130 						&  1.4			& 6				 	 &		1.25    &  		0.2   & 	1.3			&		19.4		 				   \\
66130 						&  1.4			& 6				 	 &		1.12    &  		0.2   & 	1.3			&		12.3		 				   \\
\bf 66130 				& \bf 1.9		& \bf 6   	 & \bf1.25  	& \bf	0.7   & 	\bf1.8	&		$\mathbf{31.3}$ 	 \\
\bf 66130 				& \bf 1.9		& \bf 6   	 & \bf1.25  	& \bf	0.2   & 	\bf1.3	&		$\mathbf{11.1}$ 	 \\
\bf 66130 				& \bf 2.4		& \bf 6   	 & \bf1.25  	& \bf	0.2   & 	\bf1.3	&		$\mathbf{7.5} $ 	 \\
\hline
\hline
66404 						&  1.4			& 8.6			 	 &		1.14    &  		0.2   & 	1.3			&		53.2			 				 \\
\bf 66404 				& \bf 1.9		& \bf 8.6    & \bf1.14  	& \bf	0.4   & 	\bf1.8	&		$\mathbf{77.1}$ 	 \\
\bf 66404 				& \bf 1.9		& \bf 8.6    & \bf1.14  	& \bf	0.2   & 	\bf1.3	&		$\mathbf{33}$ 	 \\
66404 						&  2.4			& 8.6			 	 &		1.14    &  		0.4   & 	1.8			&		47			 				 \\
\bf 66404 				& \bf 2.4		& \bf 8.6    & \bf1.14  	& \bf	0.2   & 	\bf1.3	&		$\mathbf{23.8}$ 	 \\
66404 						&  2.4			& 7.7			 	 &		1.08    &  		0.2   & 	1.3			&		13.7			 				 \\
\hline
\end{tabular}}
\end{table}

For the $R/L_{Ti}=6$ discharge 66130, the simulation with the nominal parameters (i.e. with the \textsc{cronos} $\hat{s}$ and $q$ values) leads to a flux value $\times\sim2.5$ above the experimental level. This discrepancy can be reduced by a reasonable variation of input parameters around the experimental uncertainties, either for $q$ and $\hat{s}$, $Z_{eff}$, or $R/L_{Ti}$. However, the discrepancy between the simulation and the experimental flux for the higher $R/L_{Ti}=8.6$ discharge - 66404 - is significantly greater. For the simulation with the base input parameters, the simulated flux is $\times\sim5$ higher than the experimental value. The simulated and experimental flux can only be reconciled by making a highly optimistic assumption with regard to the simultaneous variation of $R/L_{Ti}$, $Z_{eff}$, $\hat{s}$, $q$, and $T_e/T_i$ around their estimated error bars - as seen in the last line of table \ref{tab:summary_sec3_2}. 

\begin{figure}[htbp]
	\centering
		\includegraphics[scale=0.7]{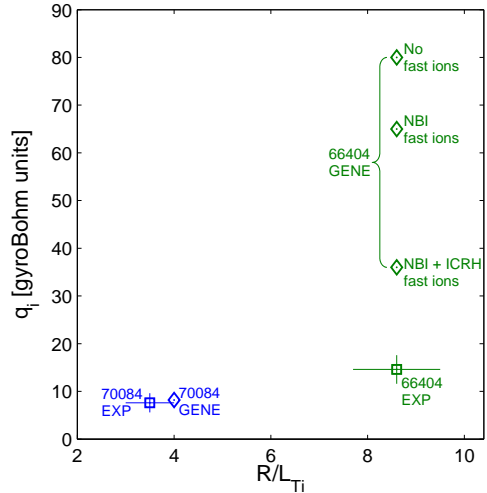}
		\caption{Comparison of measured heat flux and predicted values from electromagnetic simulations of discharge 66404 with: no fast ions (as shown in green in Fig.~\ref{fig:figure19}), NBI induced active fast ions, and both NBI + ICRH induced active fast ions (the nominal case). The 70084 measured and predicted ion heat flux is also shown for reference. The electromagnetic stabilization is enhanced by the fast ion pressure gradient and the fully nominal case is only $\times2.5$ above the experimental value.}
	\label{fig:figure20}
\end{figure}

The agreement between the simulations and measured flux values for 66404 can however be significantly improved by including the fast ion species as active species in the electromagnetic non-linear gyrokinetic simulations. The results are shown in Fig.~\ref{fig:figure20}, and are discussed in deeper detail in Ref.~\cite{citr13}. Briefly, at this radius, the fast ion induced suprathermal pressure gradient dominates the total pressure gradient, and augments significantly the electromagnetic stabilization discussed in section~\ref{subsec:beta}. With the combined contribution of the NBI and ICRH induced suprathermal pressure gradients, the calculated ion heat flux is only a factor of ${\times}2.5$  above the experimental value, which is a discrepancy that can then be explained by a reasonable variation of input parameters around the experimental uncertainties. This inclusion of active fast ion species in the electromagnetic simulations, and the subsequent stabilization of the ITG turbulence, is a key factor for reconciling the experimental observations and the simulations. 

From dedicated simulations, the fast ion stabilization was seen not to be an effective stabilising factor for the high stiffness 73221 case, showing that this mechanism can separate the high and low stiffness branches. This is due to the lower $\beta$ and lower thermal and suprathermal pressure gradients in 73221 compared with 66404, reducing significantly the impact of the electromagnetic stabilisation. Furthermore, EVE~\cite{dumo09} simulations of the ICRH power deposition profile show that the 73221 profile is significantly narrower than the 66404 case, suggesting that the 73221 suprathermal pressure profile may not overlap the experimentally relevant $\rho=0.33$ location, in line with the separation of the two branches. However, EVE presently does not include finite orbit width effects. The precise suprathermal pressure gradients could thus not be calculated with EVE. We note that the electromagnetic stabilization effect is still weak even if we assume an overlap of the 73221 suprathermal pressure profile with the $\rho=0.33$ location. %This is a highly unlikely scenario, and leads us to conclude that the stiffness in these specific gyrokinetic simulations is higher than the experimental observations for the high-rotation branch. 

%\begin{figure}[htbp]
	%\centering
	%	\includegraphics[scale=0.7]{figure_extra2.eps}
	%	\caption{}
	%\label{fig:figure_extra1}
%\end{figure}

\begin{figure}[tbp]
	\centering
		\includegraphics[scale=0.7]{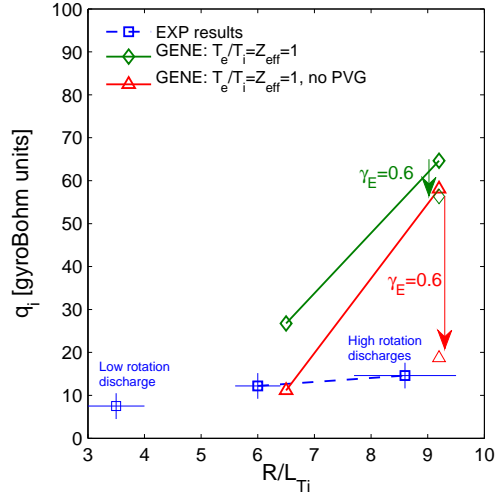}
		\caption{Comparison of flux values from non-linear simulations and experimental power balance for the high rotation discharges 66130 and 66404 at $\rho=0.33$. Here we assumed for simplicity $T_e/T_i=Z_{eff}=1$. $\hat{s}/q=0.2/1.3$ in these simulations. Sets of simulations both including and excluding the PVG drive are shown. For each set, additional 66404 simulations with $\gamma_E$ increased from 0.3 to 0.6 were carried out.}
	\label{fig:figure21}
\end{figure}

\begin{figure}[htbp]
	\centering
		\includegraphics[scale=0.7]{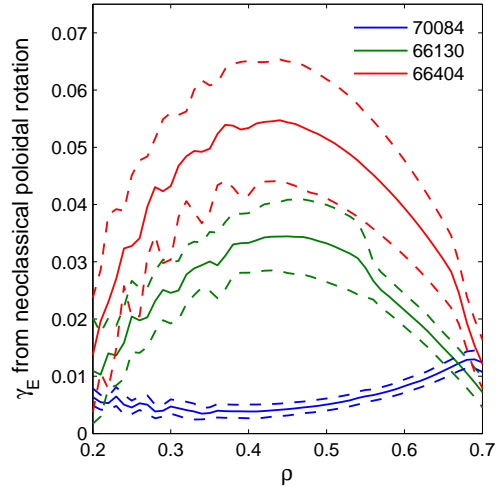}
		\caption{$\gamma_E$ derived from the \textsc{nclass} predicted poloidal rotation for deuterium. The solid lines are the average values over the 1~s time window studied for each case. The dashed lines corresponded to the standard deviation of the profiles around the mean during the time window.}
	\label{fig:figure22}
\end{figure}

The predicted and experimental fluxes for 66404 can also be reconciled by both artificially increasing $\gamma_E$ beyond the measured value from the toroidal rotation, and simultaneously ignoring PVG destabilisation. This is shown in an additional set of simulations displayed in Fig.~\ref{fig:figure21}. This assumption is consistent with assuming non-negligible poloidal rotation. However, our original assumption of negligible poloidal rotation due to neoclassical damping was justified according to \textsc{nclass}~\cite{houl97} neoclassical poloidal rotation predictions for the deuterium species within the \textsc{cronos} modelling. This is seen in Fig.~\ref{fig:figure22}, where the $\gamma_E$ profiles derived from the \textsc{nclass} predicted poloidal rotation are shown. While there is an increase in $\gamma_E$ correlated with increasing $R/L_{Ti}$ as expected, the absolute values are - while not entirely negligible for the 66130 and 66404 cases - still approximately an order of magnitude below the values necessary to provide significant turbulence suppression as observed. However, poloidal rotation values significantly above neoclassical values have been observed within internal transport barriers (ITBs)~\cite{crom05}, and \textsc{nclass} predictions have also been shown to deviate from experimentally measured carbon and main ion poloidal rotation values at DIII-D~\cite{solo06,grie12}. While we deem it unlikely that anomalous poloidal rotation is an important mechanism for flux reduction in the discharges we investigate here, in light of these observed discrepancies with neoclassical theory it is still nonetheless of interest to directly measure poloidal rotation in this class of low-stiffness-regime discharges, to examine whether any anomalous poloidal rotation is observed. It is also of interest to examine the extent to which theoretical mechanisms for generation of anomalous poloidal flow - potentially via a turbulent Reynolds stress - can play a role for cases with a high degree of external toroidal momentum injection.

\subsubsection{Summary of analysis at $\rho=0.33$}
We now summarise the entire discussion on the low-stiffness question. The predicted impact of the differences in parameters between the low rotation discharge 70084 and the high rotation discharges 66130 and 66404 at $\rho=0.33$ were examined in detail with linear and non-linear gyrokinetic simulations to investigate the potential factors leading to the observed reduced stiffness in the high-rotation cases. It was found that the differences in $R/L_n$ and the effect of rotation on the equilibrium have negligible impact on the stiffness for our parameters. The effect of rotation itself, and of the fast particle content in the high rotation cases, have non-negligible but insufficient impact to explain the observed difference in rotation. The impact of $q$ and $\hat{s}$ on the stiffness level is however significant. The non-linear stabilisation of ITG turbulence due to electromagnetic effects ($\beta_e$) was significant, reduces stiffness, and is further enhanced by including active fast ion species in the electromagnetic simulations. When self-consistently including all effects, the ion heat flux values predicted by the gyrokinetic simulations agreed with the observed values in the low rotation case (70084), and were approximately $\times2.5$ higher than the observed values for the high rotation cases 66130 and 66404. For reasonable variations of the input parameters around their uncertainties, the simulated and experimental flux values for both 66130 and 66404 could be reconciled. Improved agreement for 66404 could also be obtained by assuming non-negligible poloidal rotation, which is an unmeasured quantity for these discharges. Poloidal rotation gradients approximately an order of magnitude higher than the predicted neoclassical values would however be necessary to achieve sufficient impact. 

%% file: sect5.tex
\section{Simulated and measured ion-heat-flux comparisons at $\rho=0.64$}
\label{sec:highshear}
In the previous section, the possible factors leading to a \textit{difference} in stiffness between the low and high rotation discharges at $\rho=0.33$ was investigated. In this section we investigate the experimental observation of a \textit{lack} of stiffness reduction with rotation between the classes of discharges at $\rho=0.64$, which attained similar $R/L_{Ti}$ values, as seen in Fig.~\ref{fig:figure1}b. Non-linear simulations with \textsc{gene} of three of the discharges were performed, with parameters matching those at $\rho=0.64$. First, reduced simulations are carried out based on 70084 parameters, varying the rotation alone and examining its impact on $R/L_{Ti}$ and the stiffness. Then, full simulations are carried out - analogous to those in section \ref{subsec:all} - and the \textsc{Gene} predicted ion heat fluxes are compared with the experimental values. For all the simulations is this section, the \textsc{cronos} calculated $q$ and $\hat{s}$ values were taken for each discharge. 

\begin{figure}[htbp]
	\centering
		\includegraphics[scale=0.7]{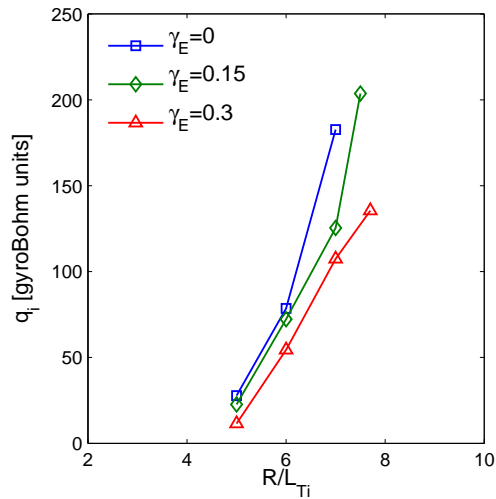}
		\caption{Non-linear $R/L_{Ti}$ scan for various levels of $\gamma_E$, based on the 70084 parameters at $\rho=0.64$. Circular geometry, $\hat{s}/q=2/3$, collisionless and electrostatic.}
	\label{fig:figure23}
\end{figure}

\begin{figure}[htbp]
	\centering
		\includegraphics[scale=0.7]{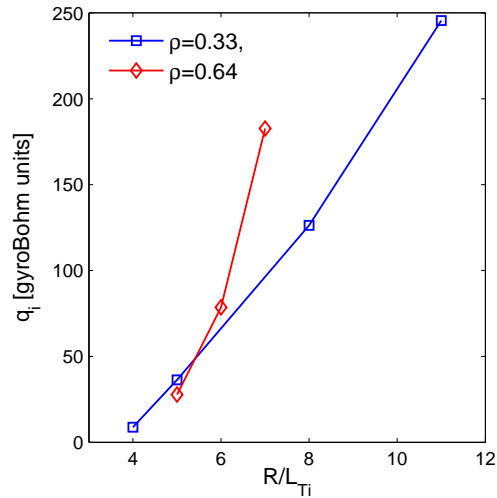}
		\caption{Non-linear $R/L_{Ti}$ scan comparing the stiffness level at $\rho=0.33$ and $\rho=0.64$, at $\gamma_E=0$, based on the 70084 parameters. Circular geometry, collisionless, and electrostatic.}
	\label{fig:figure24}
\end{figure}

In Fig.~\ref{fig:figure23} a non-linear $R/L_{Ti}$ scan with various levels of $\gamma_E$ is shown. The scan is based on discharge 70084 parameters, but uses circular geometry, $\hat{s}/q=2/3$, and is collisionless and electrostatic. The simulated stiffness is indeed greater than the $\rho=0.33$ case shown in Fig.~\ref{fig:figure5} as can be seen in a direct comparison shown in Fig.~\ref{fig:figure24} for the $\gamma_E=0$ case. Moreover, the degree of experimental $\gamma_E$ variation between the discharges (between $\gamma_E=0.1-0.3$) is also not sufficient to lead to a difference beyond typical error bars in $R/L_{Ti}$, for the same level of flux. 

Examining the differences in experimental parameters for all 3 discharges between $\rho=0.33$ and $\rho=0.64$ in table~\ref{tab:summary2}, we can see that both $\hat{s}$ and $q$ are higher at $\rho=0.64$, and $\beta_e$ is lower. All of these differences are expected to lead to higher stiffness in the $\rho=0.64$ cases compared with $\rho=0.33$. These qualitative differences in $q$-profile and $\beta_e$ between low and high radii are generic (apart from special cases such as in ITB discharges), and should hold in general in tokamak discharges. 

\begin{figure}[htbp]
	\centering
		\includegraphics[scale=0.7]{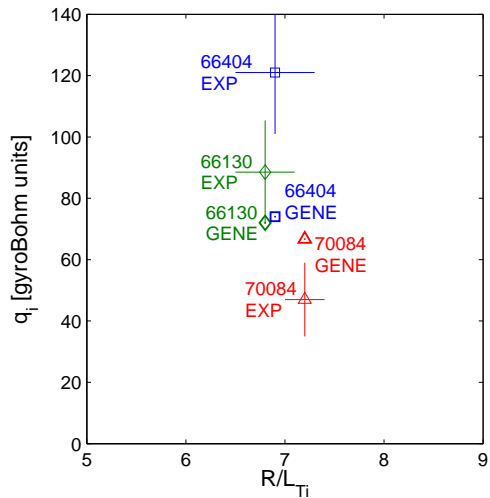}
		\caption{Comparison between gyrokinetic simulations and experiment at $\rho=0.64$ for all three discharges. The experimental values (with the error bars) are shown for 70084 (red marker), 66130 (green marker) and 66404 (blue marker). The simulated values are shown with the same colour coding and marker style for all three discharges. Runs were electromagnetic, with collisions, and with numerical geometry.} %In addition, simulations identical to the 70084 simulation but with reduced $R/L_{Ti}$ were also carried out to gain a measure of the stiffness at realistic parameters. 
	\label{fig:figure25}
\end{figure}

In Fig.~\ref{fig:figure25}, the full comparison between the simulations and the experiments is shown. These gyrokinetic simulations are electromagnetic, collisional, with numerical geometry, include a carbon species at a density consistent with $Z_{eff}=1.9$ for 66130, and $Z_{eff}=2.4$ for 70084 and 66404. The simulations include the experimental $T_e/T_i$. For all cases, the simulated and experimental ion heat flux agree approximately within $50\%$. This magnitude of difference can be easily reconciled within the reasonable uncertainties of input modelling parameters such as $R/L_{Ti}$, $T_e/T_i$ or $Z_{eff}$, particularly for these stiff transport cases. Furthermore, the far off-axis ICRH driven suprathermal pressure profile was not included in the 70084 simulation, which may explain a proportion of the overprediction observed. An $R/L_{Te}$ sensitivity check for discharge 66130 was carried out, which had the largest relative $R/L_{Te}$ error throughout the data set, as seen in table~\ref{tab:summary2}. It was found from the dedicated non-linear simulations that within the possible $R/L_{Te}$ range the impact on ion transport is minimal. %For example, an additional 66404 simulation carried out with $Z_{eff}=2.4$, $T_e/T_i=1.1$ and $R/L_{Ti}=7.8$ led to agreement with the experimental flux within the experimental error-bars. 

To summarise, the effect of rotation alone at $\rho=0.64$ is not expected to lead to experimentally discernible differences in $R/L_{Ti}$ and stiffness for the range of experimental $\gamma_E$ examined. This is in agreement with the experimental trend seen in Fig.~\ref{fig:figure1}b. When comparing the full non-linear gyrokinetic ion heat flux predictions with the experimental values at $\rho=0.64$, general agreement within reasonable input parameter uncertainties is seen for all the discharges, both at high and low rotation.

%% file: sect6.tex
\section{Conclusions}
\label{sec:discuss}

Observations at JET have shown evidence of reduced ion temperature profile stiffness correlated with low magnetic shear and increased flow shear. The same data-set has also raised questions regarding the experimental validation of the Dimits shift paradigm, and the low-rotation subset of discharges within this data-set seemed to display \textit{higher} profile stiffness than expected from gyrokinetic simulations. These observations have motivated extensive non-linear gyrokinetic simulations to investigate these questions. Simulations using the \textsc{Gene} code were carried out, with parameters based on a subset of these JET discharges. Transport sensitivity scans of various parameters that differed between the discharges - aside from rotation - were carried out, to assess potential mechanisms that may explain the observations. Full simulations including electromagnetic effects, numerical geometry, $Z_{eff}$, experimental $T_e/T_i$, and rotation were also performed at $\rho=0.33$ (in the low stiffness zone) and $\rho=0.64$ for the discharges studied. The predictions were compared with the experimental results. The conclusions can be summarised as follows:

\begin{enumerate}[(1)]
\item The transport sensitivity to $R/L_n$ variations, dilution due to fast particles, increased Shafranov shift due to suprathermal pressure, and the effect of rotation on the equilibrium, were all examined. It was established that none of the above factors are sole mechanisms for the transition to the reduced stiffness regime. Their cumulative effect is however not negligible - particularly that of fast particles both through dilution and an increased Shafranov shift.
\item The sensitivity of the transport to $\beta_e$ was examined. It was established that even for the relatively low $\beta_e$ values present in these discharges, the non-linear electromagnetic ITG stabilisation is significant. This stabilisation, at least for $\beta_e<0.48\%$, is a stiffness reduction as opposed to a threshold shift for discharge 66404. The non-linear stabilisation is significantly greater than the linear $\beta_e$ stabilisation, and may be related to an increased relative amplitude of zonal modes. The effect is further enhanced by the addition of active fast ion species in the electromagnetic simulations, whose pressure gradients add to the electromagnetic coupling. Further investigations of the parameterisation of this effect is important for incorporation into the `mixing length rule' of quasilinear transport formulations. It is expected that this effect would, using such formulations, lead to more optimistic predictions for the energy confinement in future devices such as ITER and DEMO, which are not expected to have significant rotation but could still benefit from electromagnetic stabilisation. particularly due to the significant fast ion content in burning plasmas.
\item No clear disagreement is observed between the experimentally observed turbulence $R/L_{Ti}$ threshold and the upshifted (Dimits shift) non-linear threshold predicted by the gyrokinetic simulations. Previously reported results of such a disagreement in Refs.~\cite{mant09,mant11} were found to be highly sensitive to the precise choice of $q$ values used for the simulations. Recently improved data processing methodology has led to a revised $q$ value now seemingly pointing to good agreement between the experimentally observed and simulated threshold values. However, a firm conclusion on this point is not justified considering the sensitivity of the results to both $q$ and $\hat{s}$.
\item For the nominal parameters for both the low and high rotation cases at $\rho=0.33$, agreement between the gyrokinetic simulations and the experimental ion heat fluxes could be obtained within reasonable variations of the input parameters within their uncertainties. While the competition between parallel velocity gradient (PVG) destabilisation and $E{\times}B$ stabilisation can reduce the stiffness in the vicinity of the turbulence threshold, the predicted flux levels themselves are still significantly higher than the experimental values. The key factor for improved agreement for the high $R/L_{Ti}=8.6$ case is obtained by electromagnetic stabilisation, enhanced by the suprathermal pressure gradients. Thus, we conclude that electromagnetic stabilisation enhanced by fast ions is the primary factor responsible for the low stiffness regime, and not rotational flow shear. Since flow shear and fast ion content is typically coupled in NBI driven discharges, it is important to devise future experiments that actively decouple these effects for further investigation.
\item For the low-rotation branch at $\rho=0.33$ within the data-set studied, the observation of seemingly anomalous high stiffness compared with the gyrokinetic simulations is likely explained by a downshift in the ITG critical gradient due to higher $T_e/T_i$ in the high flux cases. However, a firm conclusion in this regard is precluded by the high sensitivity of the critical gradient to $q$ and $\hat{s}$, and thus to the $q$-profile uncertainties. 
\item The gyrokinetic predictions and experimental fluxes were also compared at $\rho=0.64$ for the three discharges. The experimental variation in flow shear between the discharges was not predicted to be sufficient to lead to a discernible difference in $R/L_{Ti}$ - in agreement with the observations. The simulated and experimental ion heat fluxes for all examined discharges all agreed to within approximately $50\%$. This degree of discrepancy can be explained by reasonable variations of the input parameters within the experimental uncertainties.  
\end{enumerate}

%% file: sect7.tex
\section{Acknowledgements}
This work, supported by the European Communities under the contract of Association between EURATOM/FOM and EURATOM/CEA, was carried out within the framework of the European Fusion Programme with financial support from NWO. The views and opinions expressed herein do not necessarily reflect those of the European Commission. This work is supported by NWO-RFBR Centre-of-Excellence on Fusion Physics and Technology (Grant nr. 047.018.002). The authors would like to thank: H. Doerk, D.R. Hatch, M. Barnes, A. Schekochihin, E. Highcock, F. Millitello, J. Garcia, M. Schneider, and N. Hawkes for stimulating discussions. The authors are extremely grateful to D.R. Mikkelsen for aiding with computational resources. This research used resources of the HPC-FF in Juelich, and the National Research Scientific Computing Center, which is supported by the Office of Science of the U.S. Department of Energy under Contract No. DE-AC02-05CH11231.